\documentclass[aps,prx,longbibliography,twocolumn,superscriptaddress]{revtex4-2}

\usepackage{amsfonts}
\usepackage{amsmath}
\usepackage{graphicx}
\usepackage{dsfont}
\usepackage{diagbox}
\usepackage{array}
\usepackage{amssymb}
\usepackage{bbm}
\usepackage{float}
\usepackage{subfigure}
\usepackage{MnSymbol}
\usepackage{url}
\usepackage{times}
\usepackage{manfnt}
\usepackage{booktabs}
\usepackage{xcolor}
\definecolor{darkblue}{HTML}{004D6B}
\definecolor{darkred}{HTML}{8c1515}
\definecolor{darkgreen}{HTML}{006400}
\usepackage{hyperref}
\hypersetup{
    pdftitle={NishimoriCatQubitLoss},
	colorlinks=true,
	urlcolor=darkred,
	citecolor=darkblue,
	linkcolor=darkred,
	breaklinks
}
\pagecolor{white}
\usepackage{ulem}
\usepackage{physics}
\usepackage{enumitem}
\usepackage{wasysym}
\usepackage{tikz}
\usetikzlibrary{arrows.meta}

\newcommand{\mc}[1]{\multicolumn{1}{c}{#1}}
\newcommand{\pmin}{\phantom{-}}
\newcommand\redsout{\bgroup\markoverwith{\textcolor{red}{\rule[0.5ex]{2pt}{0.4pt}}}\ULon}

\begin{document}

\title{Flow to Nishimori universality in weakly monitored quantum circuits with qubit loss} 

 \author{Malte Pütz}
 \affiliation{Institute for Theoretical Physics, University of Cologne, Z\"ulpicher Straße 77, 50937 Cologne, Germany}

\author{Romain Vasseur}
\affiliation{Department of Theoretical Physics, University of Geneva, 24 quai Ernest-Ansermet, 1211 Gen\`eve, Switzerland}

\author{Andreas W.W. Ludwig}
\affiliation{Department of Physics, University of California, Santa Barbara, California 93106, USA}

 \author{Simon Trebst}
 \affiliation{Institute for Theoretical Physics, University of Cologne, Z\"ulpicher Straße 77, 50937 Cologne, Germany}

 \author{Guo-Yi Zhu}
\email{guoyizhu@hkust-gz.edu.cn}
\affiliation{The Hong Kong University of Science and Technology (Guangzhou), Nansha, Guangzhou, 511400, Guangdong, China}

\date{\today}
\begin{abstract}
In circuit-based quantum state preparation, qubit loss and coherent errors 
are circuit imperfections that imperil the formation of long-range entanglement beyond a certain threshold.
The critical theory at the threshold is a continuous
entanglement transition known to be described by a
(2+0)-dimensional non-unitary conformal field theory which, for the two types of imperfections of certain circuits, 
is described by either percolation or Nishimori criticality, respectively.  
Here we study the threshold behavior when the two types of errors {\sl simultaneously} occur
and show that, when moving away from the Clifford-regime of projective stabilizer measurements, 
the percolation critical point becomes unstable and the critical theory flows to Nishimori universality.
We track this critical renormalization group (RG) crossover flow by mapping out the entanglement phase diagrams, parametrized by the probability and strength  of random weak measurements, of two dual protocols preparing surface code or GHZ-class cat states from a parent cluster state via constant-depth circuits. 
Extensive numerical simulations, using hybrid Gaussian fermion and tensor network / Monte Carlo sampling techniques on systems with more than a million qubits,
demonstrate that an infinitesimal deviation from the Clifford regime leads to 
 a sudden, strongly non-monotonic
entanglement growth at the incipient non-unitary RG flow.
We argue that spectra of scaling dimensions of both the percolation and Nishimori fixed points  exhibit {\it multifractality}. 
For percolation, we provide
the exact (non-quadratic) multifractal spectrum of exponents,
while for the Nishimori fixed point we show high-precision
numerical results for
five leading exponents characterizing multifractality.
\end{abstract}

\maketitle

%%%%%%%%%%%%%%%%%%%%%%%%%%%%%%%%%%%%%%%%%%%%%%%%%%%%%%%%%%%%%%%%%%%%%%%%%%%%
% Introduction
%%%%%%%%%%%%%%%%%%%%%%%%%%%%%%%%%%%%%%%%%%%%%%%%%%%%%%%%%%%%%%%%%%%%%%%%%%%%

Despite the fragility of individual qubits to the effects of leakage, measurement and decoherence, quantum information stored in many physical qubits 
can be remarkably robust.
The surface code, in which long-range many-qubit entanglement protects a single logical qubit, is the quintessential embodiment of this idea and has become a principal design element for fault-tolerant quantum computing architectures. In quantitative terms, the robustness is given by a threshold value that often separates a ``correctable" from an ``incorrectable" regime and which has been determined for various types of incoherent~\cite{Preskill2002} and coherent~\cite{teleportcode} noise channels 
as well as qubit loss~\cite{Grassl97erasure, Stace2009, scholl2023erasure, teoh2023dualrail}.  
These thresholds, however, extend beyond the realm of quantum error correction codes and, taking the example of qubit loss (also known as erasure, leakage or heralded noise), 
can also be studied to discuss the stability of measurement-based quantum computation~\cite{Browne2008}, photonic quantum computing~\cite{pant2019percolation}, 
or open quantum dynamics~\cite{Burnell24heralded}. 
But while the numerical value of the threshold might vary for these various settings, they share the commonality that the underlying noisy protocols often fall into the Clifford~\cite{Gottesman97} regime,
which allows for an efficient classical numerical computation of the critical threshold. The critical theory at the threshold, however, will depend on the specifics of the noise channel.
For qubit loss, the fundamental phase transition is intrinsically linked to the {\sl percolation problem},
a staple of classical statistical mechanics \cite{Stauffer1992}. 
From a field theoretical perspective, the percolation transition is, in a sense, the simplest, best-understood, and also analytically most tractable
example of disorder-induced criticality described by a non-unitary conformal field theory (CFT)~\cite{GURARIE1993, Ludwig05c0log, GURARIE1999, Vasseur_2012, Cardy2013log}. 

In recent times, with interest shifting to critical phenomena in entangled states of matter subject to measurement-induced randomness and 
non-equilibrium phase transitions in monitored quantum systems~\cite{Fisher2022reviewMIPT,Potter21review},
percolation again serves not only as the simplest instance of such phenomena~\cite{Nahum20freefermion,Lang_2020,Vasseur24transfer}, 
but also describes~\cite{Jian2020,Altman2020weak} 
the tractable limit (at infinite on-site Hilbert space dimension) of such a generic transition.
In this context, percolation criticality also naturally arises in the context of measurement-only Clifford systems~\cite{Nahum20freefermion,Lang_2020,Vasseur24transfer,Ludwig24stat}, which are generated by {\it projective} measurements of the Pauli string operators~\cite{Gottesman97}.
As such, it represents one of the simplest examples in which to  study the question of  effects of {\it non-Clifford} perturbations of monitored stabilizer quantum systems.
These questions have been the driving motivation for the study reported in this manuscript.

In a quantum circuit, non-Clifford perturbations of even the smallest scale (such as, say, an imperfect $\pi/2$ qubit rotation~\footnote{
Clifford perturbations are restricted to a discrete set of operations. Take single-qubit as an example, the Clifford perturbations must rotate the qubit in the Bloch sphere by some 90 degree angle, such that it rotates between the axis $X$ or $Y$ or $Z$ that forms the octahedron. A rotation by a generic angle (e.g. the parameter $2t$ later introduced in our paper where $t\neq \pi/4$.) that takes the qubit off the octahedron to the more general direction in the Bloch sphere is beyond Clifford.
}) drastically change the system -- they immediately 
allow the system's wavefunction to spread into a broader Hilbert space, 
while at the same time leading to an explicit violation of the Gottesman-Knill condition for classical simulability~\cite{Gottesman97}.
Due to these fundamental changes caused by non-Clifford perturbations, the latter are clearly expected to induce a change in critical behavior, a fact that is well established in the context of generic quantum circuits in the
above mentioned
limit of large on-site Hilbert space 
dimension~\cite{Jian2020}. While the latter example provides a `proof of concept', in the present work we are interested in more `practical' manifestations of these effects.
In fact, all of the currently available quantum circuit platforms do not implement pristine Clifford gates but exhibit a (small) level of imperfections (lending to their classification as noisy intermediate-scale quantum (NISQ)~\cite{Preskill2018nisq} devices), making it a pressing question whether the threshold behavior they exhibit is dominated by percolation or some other universality class(es).
   
\begin{figure}[tb] 
   \centering
   \includegraphics[width=\columnwidth]{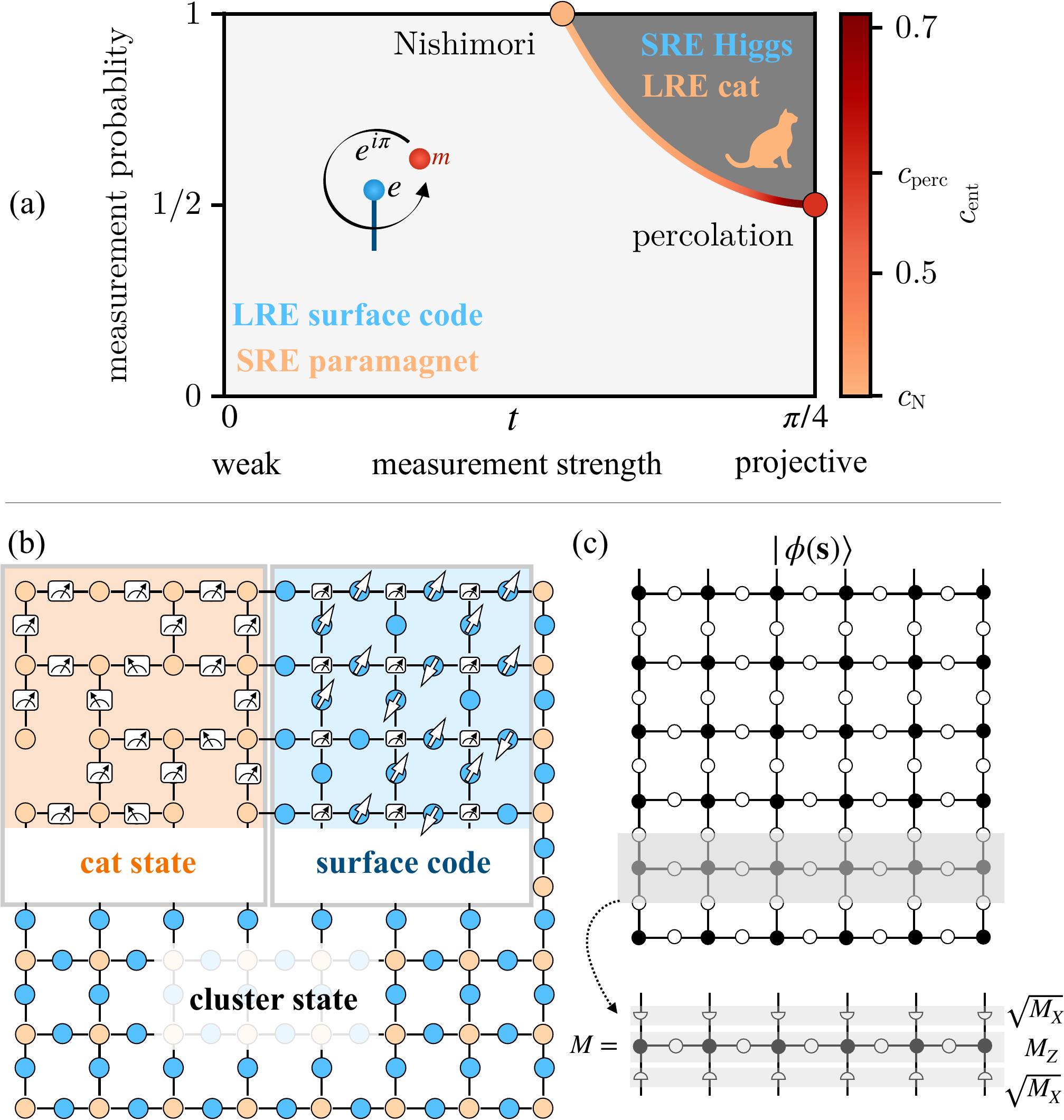} 
   \caption{{\bf Entanglement phase diagram, protocols and tensor network}. 
  (a) Long-range entanglement phase diagram of GHZ-class cat state {\it or} topological surface code, driven by random and weak measurements. 
  Here ``LRE'' stands for long-range entangled phase, and ``SRE'' stands for short-range entangled phase. The defining character of LRE surface code phase is the anyon excitations $e$ and $m$ particles which pick up a $\pi$ phase when winding around each other~\cite{Kitaev2003}. The condensation of $e$ particles leads to the SRE ``Higgs'' phase~\cite{Zhu19}. Wegner's duality relates LRE surface code to SRE paramagnet, and SRE Higgs phase to a LRE cat phase. 
   (b)
   Protocol starting from a parent cluster state. 
   With respect to a square lattice, the orange spheres are denoted as ``site'' qubits, while the blue spheres are denoted as ``bond'' qubits. Together, they form a 2D Lieb lattice. The first protocol projectively measures out the bond qubits in a rotated basis, which effectively induces two-body weak parity measurements for the adjacent site qubits (orange spheres): $\exp{-\beta s_{ij} Z_i Z_j/2}$, resulting in a cast sate for the site qubits~\cite{NishimoriCat}. 
   For the second protocol, projectively measuring out the site qubits results in a surface code for the bond qubits (blue spheres), which can be further subject to weak measurements $\exp{\beta s_{ij} Z_{ij}/2}$ with a probability~\cite{teleportcode}, denoted by the white hollow arrow in the top right ``surface code'' lattice of panel (b).
   (c) Tensor network, sliced into a sequence of transfer matrix product operators. Each circle on the bond denotes a matrix, which is chosen to capture the Ising interaction or null interaction depending on the weak or null measurement. Smooth boundary condition is shown, but rough boundary condition for surface code can be implemented by sending the left and right boundary white circles to be identity matrices.
   }
    \label{fig:schematic}
\end{figure}

In this manuscript, we study the effect of non-Clifford perturbations on the percolation transition induced by qubit loss by adding a second noise channel
that is known to give rise to distinct critical behavior -- Nishimori physics, induced by weak (stabilizer) measurements that give rise to a (tunable) coherent error \cite{teleportcode,NishimoriCat,Lee22,Chen25nishimori}. 
Putting these two noise channels into direct competition yields a two-dimensional phase diagram where the two critical theories are connected
via a line of thresholds, as depicted in Fig.~\ref{fig:schematic}. Our key result is that, upon moving away from the Clifford limit by introducing infinitesimal coherent noise, an RG flow sets in and percolation flows to Nishimori criticality, making the latter the universal fixed point in the presence of both noise channels. We demonstrate this flow by calculating a number of characteristic properties of the critical theories along the threshold line, including central charge estimates, critical exponents, along with various cumulants. 

%%%%%%%%%%%%%%%%%%%%%%%%%%%%%%%%%%%%%%%%
%%%%%%%%%%%%%%%%%%%%%%%%%%%%%%%%%%%%%%%%

\subsection*{Nishimori criticality in monitored quantum systems}
Here we briefly review the criticality of the Nishimori and percolation transitions emerging in the monitored quantum circuits. 
We start from qubits array on a Lieb lattice (Fig.~\ref{fig:schematic}(b)), which can be viewed as a square lattice with a set of qubits on the sites and a set of qubits on the bonds. By entangling a bond qubit with its adjacent two qubits at sites $i$ and $j$, and then projectively measuring out the middle bond qubit, one can achieve a two-body parity measurement for the two site qubits with tunable strength~\cite{NishimoriCat}:
\begin{equation}
e^{\frac{\beta}{2} s_{ij} Z_i Z_j} \propto e^{\frac{\beta}{2}} \frac{1+s_{ij} Z_i Z_j}{2}+e^{-\frac{\beta}{2}}\frac{1-s_{ij} Z_i Z_j}{2} \ ,
\label{eq:weakmeasurement}
\end{equation}
where $\beta=\tanh^{-1}\sin(2t)$ can be controlled by an entangling gate parameter $t\in[0,\pi/4]$, and $s=\pm 1$ is the binary measurement outcome of the bond qubit. 
$t=\pi/4$ corresponds to a maximally entangling gate between the site qubits and the bond qubit~\footnote{Specifically, the entangling gate between the site qubit whose Pauli Z matrix is denoted by $Z_j$, and the bond qubit whose Pauli Z is denoted by $Z_{ij}$, is an RZZ rotation gate $\exp{-i t Z_i Z_{ij}}$ with evolution time $t$, see Ref.~\cite{NishimoriCat, Chen25nishimori} for more details.}, which realizes the strongest, projective measurements at $\beta=\infty$. This can be equivalently realized in a cluster state on Lieb lattice~\cite{Verresen2021cat,NishimoriCat, JYLee}. Due to the randomness of the measurement outcomes, the projective measurement limit $\beta=\infty$ results in a GHZ state~\cite{GHZ} subjected to random spin flips conditioned on the measurement outcome configuration $\{s_{ij}\}$. When $\beta<\infty$, the random 2-dimensional state, cast as a projected entangled pair state (PEPS)~\cite{Cirac06}, can be mapped to a random bond Ising model (RBIM) on the square lattice~\cite{NishimoriCat}. 
It was found that Born's rule equates the probability of the distribution $\{s_{ij}\}$ to the effective partition function of the ensuing RBIM~\cite{NishimoriCat}. This satisfies a key defining feature of the Nishimori line in the phase diagram of the classical RBIM - namely, the disorder probability after a gauge symmetrization is exactly equal to the thermal partition function~\cite{Nishimori1981}. 
Thus Born's rule rigorously forces such monitored quantum system to trace the Nishimori line through the classical phase diagram. And the phase transition falls into the Nishimori universality class. Such robustness of the Nishimori criticality can be shown~\cite{Puetz25learning} to survive finite measurement readout noise, which is verified in a quantum experiment~\cite{Chen25nishimori}.
In the following, we introduce qubit loss error to the aforementioned protocol to investigate the competition between the two types of errors driving the two representative critical theories in monitored quantum circuits. Moreover, we go beyond the measurement based state preparation of the GHZ state and consider its duality counterpart - the topological surface code under errors, which can be found to be captured by exactly the same phase diagram and criticality.

\subsection*{Quantum circuit}

Our two principal circuits, illustrated in Fig.~\ref{fig:schematic}(b), are shallow (constant-depth) monitored quantum circuits that result in 2D PEPS for qubits defined on the sites, or the bond centers of a square lattice, respectively. The collection of the sites and the bond centers forms a 2D Lieb lattice, which provides a unifying perspective. The cluster state on such Lieb lattice can be viewed as a parent state for both protocols. Here for our convenience, the cluster state is defined in detail as follows. Denoting the site of a square lattice by $i$ and $j$, and nearest neighbouring (in square lattice) pair by $\langle ij\rangle$, then a qubit on the bond center is labeled by $Z_{ij}$. Our cluster state is the eigenstate of the following local interaction terms: $Z_i Z_{ij} Z_j = +1$ for any bond $\langle ij \rangle$ and $X_j \prod_{\langle ij\rangle}X_{ij} = +1$ for any site $j$ of the square lattice. 

For the first protocol, we start from the cluster state and projectively measure the bond qubits in a rotated basis~\cite{JYLee}, which leaves an ensemble of 2D PEPS on the site qubits. In view of the site qubits, the cluster state preparation and the projective bond qubits measurements can be combined as a method to achieve the two-body weak parity measurements of tunable strength as in Eq.~\eqref{eq:weakmeasurement}~\cite{NishimoriCat, Chen25nishimori}, resulting in GHZ-class cat states \cite{Raussendorf2001}. 
This can be generalized to generic case where the Lieb lattice has extensive bond vacancies. The resulting random 2D PEPS state can be compactly written as 
\begin{equation}
\begin{split}
&\ket{\psi(\mathbf{s})}_{\text{cat}} \propto   \exp\left(\frac{\beta}{2}\sum_{ij\in\mathcal{C}} s_{ij} Z_i Z_j\right)\ket{+}  \ ,\\
\end{split}
\label{eq:psi}
\end{equation}
where $j$ denotes sites of the square lattice, 
$s_{ij}=\pm1$ is the measurement outcome on a bond, ${\bf s} := \{s_{ij}\}$ denotes the set of all measurement outcomes,
and $\mathcal{C}$ denotes the set of bonds that are measured. 
Later we will use $N$ to denote the total number of sites, and $ N_M =N_B \cdot p_{\text{meas}}$ is the total number of measured bonds that are chosen randomly with measurement probability $p_{\text{meas}}$. 

For the second protocol, we start from the cluster state and first projectively measure the site qubits in $X$ basis~\cite{Raussendorf2005}. Since $Z_i Z_j Z_{ij}=+1$, this effectively performs a Wegner's duality~\cite{Wegner71duality} to map the domain wall
degree of freedom $Z_i Z_j=\pm 1$
on a bond $\langle ij \rangle$ to a spin residing on the bond center $Z_{ij}=\pm 1$~\cite{Nat2021measure}, up to randomness that can be corrected~\cite{Raussendorf2005}. 
Thus the paramagnet $\ket{+}$ for qubits on sites with proliferated domain walls is mapped to the toric code state $\ket{\psi_{\rm TC}}$ for qubits on the bond centers, as a loop condensate~\cite{Kitaev2003}:
\begin{equation}
|\psi_{\rm TC}\rangle = 
\prod_{\square}\frac{1+\prod_{\langle ij\rangle\in \square} Z_{ij}}{2} |+\rangle^{\otimes N_B} \ ,
\end{equation}
where $N_B$ is the total number of bond qubits and $\square$ denotes a square plaquette of the square lattice. One can check that it satisfies $\prod_{\langle ij\rangle \in \square} Z_{ij}\ket{\psi_{\rm TC}} = \ket{\psi_{\rm TC}}$ and $\prod_{\langle ij\rangle \in +}X_{ij}\ket{\psi_{\rm TC}} = \ket{\psi_{\rm TC}}$ where $+$ denotes a star vertex of the square lattice, following the convention of Ref.~\cite{Kitaev2003}.
Applying Wegner's duality beyond the fixed point of $\beta=\infty$, we can dualize the whole random PEPS Eq.~\eqref{eq:psi} to 
\begin{equation}
\begin{split}
&\ket{\psi(\mathbf{s})}_{\text{code}} \propto   \exp\left(\frac{\beta}{2} \sum_{ij\in\mathcal{C}}s_{ij} Z_{ij}\right)\ket{\psi_{\rm TC}} \ .
\end{split}
\label{eq:psicode}
\end{equation}
To realize this ensemble of wave functions, we need to further perform (non-projective) weak measurements with a tunable strength $\beta$ for the bond qubits.

Comparing the two protocols, 
$\ket{\psi(\mathbf{s})}_{\text{cat}}$ are cat states prepared by {\it weak} parity measurements~\cite{NishimoriCat} over randomly chosen bonds of a square lattice, where the nonuniform measurement can be attributed to the loss of ancilla qubits. $\ket{\psi(\mathbf{s})}_{\text{code}}$ are surface code states subject to {\it weak measurements} for randomly chosen physical qubits~\cite{teleportcode}. The finite probability of weak measurements can arise from the auxiliary qubit loss or erasure errors. 

The two protocols  remain {\it dual} to each other in the presence of errors and share the same phase transition.
To see this, let us expand the 2D wave function~\eqref{eq:psi} in the computation basis: $\sum_\sigma \exp\left(\beta/2 \sum_{\langle ij\rangle \in \mathcal{C}} s_{ij} \sigma_i\sigma_j\right) \cdot |\{\sigma_i\}\rangle$, where the wave function amplitude can be interpreted as square root of Bolztmann weight of a classical statistical model, akin to a Rokhsar-Kivelson (RK) type of state~\cite{Henley04RK,Fradkin04RK, Isakov_2011}. 
Consequently, Born's rule for measurement outcomes allows us to interpret the probability distribution function of the state in one-to-one correspondence 
to the partition function of the classical statistical model,
\begin{equation}
P(\mathbf{s})=\braket{\psi(\mathbf{s})}  \propto \sum_{\sigma} \exp\left( \beta \sum_{\langle ij\rangle \in \mathcal{C}} s_{ij} \sigma_i\sigma_j\right) \ ,
\label{eq:P}
\end{equation}
where $\mathcal{C}$ denotes the  set of measured bonds, i.e.\ the 
geometric connected~\footnote{We need to be above the percolation threshold, $p_{\rm meas}\geq 1/2$.}
cluster on which 
the effective statistical model is supported -- this is an incarnation of the random bond Ising model (RBIM) with 
random bond dilution~(see, e.g., Ref.~\onlinecite{Picco2006}).
For $p_{\text{meas}}=1$ (i.e.\ when all bonds are being subject to measurements) this model reduces to the standard RBIM on a square lattice~\cite{NishimoriCat, JYLee}, 
with $P(\mathbf{s})$ capturing the partition function 
in the presence of disorder realizations $\mathbf{s}$ whose characteristics are tuned by the measurement strength $0 \leq t \leq \pi/4$, 
with a Nishimori transition occurring at intermediate strength  $t_c\approx 0.143\pi$.
In the projective measurement limit $t=\pi/4$, where the circuit falls into the Clifford class~\cite{Stace2009, Lang_2020, Vasseur24transfer, Yoshida24monitoredcode}, 
we can sweep $0 \leq p_{\text{meas}} \leq 1$ to deplete the number of measured bonds, which induces a percolation transition at $p_{\text{meas}} = 1/2$ on the square lattice. 

Beyond these two limits, we obtain the  numerically computed global phase diagram for arbitrary combinations $(t, p_{\text{meas}})$ of measurement strength and bond dilution 
as shown in Fig.~\ref{fig:schematic}(a). Our focus will be on the critical line connecting the two critical points introduced above.

%%%%%%%%%%%%%%%%%%%%%%%%%%%%%%%%%%%%%%%%%%%%%%%%%%%%%%%%%%%%%%%%%%%%%%%%%%%%
% Tensor network perspective
%%%%%%%%%%%%%%%%%%%%%%%%%%%%%%%%%%%%%%%%%%%%%%%%%%%%%%%%%%%%%%%%%%%%%%%%%%%%
\subsubsection*{Equivalent (1+1)D monitored circuit dynamics}
One useful alternative perspective is to view the 2D quantum problem as a (1+1)D quantum problem using the transfer matrix.
The transfer matrix for the evolution at a given time slice denoted by $y$ %$t$ 
can 
be decomposed as $M(y) = \sqrt{M_X} \cdot M_Z(y) \cdot \sqrt{M_X}$, where $M_Z(y)$ and $M_X$ are the Ising interaction evolution 
and transverse field evolution gates
\begin{equation}
   \begin{split}
	M_Z(y) = \exp\left(\beta  \sum_{x=1}^{L_x} s_x(y) Z_x Z_{x+1}\right) \ ,\ 
  	M_X =  \exp\left( \beta' \sum_{x=1}^{L_x} X_x\right) \ ,
   \end{split}
   \end{equation}
where $s_x(y)=\pm 1$ is the bond disorder at site
$x$ and time $y$, after a gauge transformation that fixes the coupling sign of the temporal links, see Appendix~\ref{sec:gaugetransform}. $\beta'=-\frac{1}{2}\ln \tanh(\beta)$ is the Kramers-Wannier dual of $\beta$. This (1+1)D quantum model is an {\it imaginary-time evolving kicked Ising chain} with bond randomness in the Ising interaction that fluctuates in space and time. 
The probability $P({\bf s})$, Eq.~(\ref{eq:P}),
which equals the partition function of the classical 2D Ising model, reading
\begin{eqnarray}
\label{LabelEqDefPhis}
P({\bf s})=
\langle +
|\phi({\bf s})\rangle, 
\ {\rm where} \ 
|\phi({\bf s})\rangle :=
\prod_{y=1}^{L_y}
M(y) \ 
|+ \rangle, 
\end{eqnarray}
(assuming the same boundary conditions at $y=1$ and $y=L_y$ in Eq.~(\ref{eq:P}), which were left unspecified in the latter equation).
Since the physical qubits of the 2D bulk
are traced out, such a 1D quantum state 
$|\phi({\bf s})\rangle $
describes the complexity of the 2D
bulk~\cite{Harrow2022}, and is related to the diagonal elements of the (mixed state)
boundary density matrix~\cite{Wang25selfdual}. 
In the Clifford limit $t=\pi/4$
(projective measurement~\cite{Yoshida24monitoredcode}), the effective (1+1)D circuit with 
random bond dilution is equivalent to the 
projective measurement-only 
Ising model, whose phase transition has been demonstrated to fall into the percolation universality class~\cite{Nahum20freefermion,Lang_2020, Vasseur24transfer}. 
In the non-diluted limit $p_{\text{meas}}=1$, the  resulting (1+1)D circuit is a 
1D deep measurement-only quantum circuit representation of the 2D RBIM (which was discussed in the statistically space-time dual  version of the same transfer matrix description as above in Ref.~[\onlinecite{Read2001}]),
which undergoes a Nishimori transition at $t_c\approx 0.143\pi$ (along the Nishimori line). 

Our model thus generalizes the projective Ising model to weak measurements and beyond the Clifford regime. We will show in the following that the percolation criticality of the Clifford limit is a singular point and will be immediately overwritten by the Nishimori universality class upon introducing non-Clifford elements. 

%%%%%%%%%%%%%%%%%%%%%%%%%%%%%%%%%%%%%%%%%%%%%%%%%%%%%%%%%%%%%%%%%%%%%%%%%%%%
% Numerical simulations
%%%%%%%%%%%%%%%%%%%%%%%%%%%%%%%%%%%%%%%%%%%%%%%%%%%%%%%%%%%%%%%%%%%%%%%%%%%%
\subsubsection*{Numerical simulations}

At the core of our numerical approaches, we generate a set
of measurement outcomes ${\bf s}$ by uncorrelated sampling followed by gauge symmetrization, as described in Ref.~\cite{NishimoriCat}. 
Compared with the relatively modest sizes $L\sim \mathcal{O}(10)$ of Ref.~\cite{NishimoriCat}, here we can collect data of much bigger system sizes $L\sim \mathcal{O}(10^3)$ because the whole phase diagram of interest has a {\it gauge symmetry} (see Appendix) that allows us to simplify the sampling of the measurement outcomes by means of gauge fixing.  
Using these sampled configurations, we employ three numerical methods to analyze the entanglement properties and phase transitions in our model.

The most versatile of these is the {\it tensor network} contraction method~\cite{NishimoriCat, Zhang22tnsupremacy, Harrow2022}, which enables us to compute the coherent information $I_c$ and the entanglement entropy $S$ across the full phase diagram. This approach allows us to access system sizes up to $(L + 1) \times L$ with linear system sizes as large as $L = 512$ 
for $I_c$, and up to $L \times 2L$ with $L=1024$ for $S$ on long stripe geometries (open boundary conditions). 
For each fixed disorder realization, the qubit-based tensor network can be fermionized into a Gaussian fermion tensor network akin to the Chalker-Coddington network model~\cite{Wang25selfdual, Jian22network,Chalker2002} which allows efficient contraction in periodic boundary conditions without truncation, which we use for the multifractality calculations below.

To target specific critical points in periodic boundary conditions, we also utilize {\it Clifford circuit} simulations at the percolation point and a {\it Majorana fermion evolution} technique at the Nishimori point. Both methods are used to compute $S$ on long cylinder geometries: $L \times 4L$ with $L=1024$ for the Clifford simulations, and $L \times 20L$ with $L=1024$ for the Majorana approach. In both cases, 
the 1D quantum state undergoing the monitored quantum dynamics reaches the steady state for $4L$ time steps prior to calculating the entanglement.
For the Majorana simulations, we compute $S$ every $L/2$ time steps after thermalization, reducing computational cost while ensuring sufficient decorrelation between samples.

%%%%%%%%%%%%%%%%%%%%%%%%%%%%%%%%%%%%%%%%
\begin{figure}[t]
   \centering
   \input{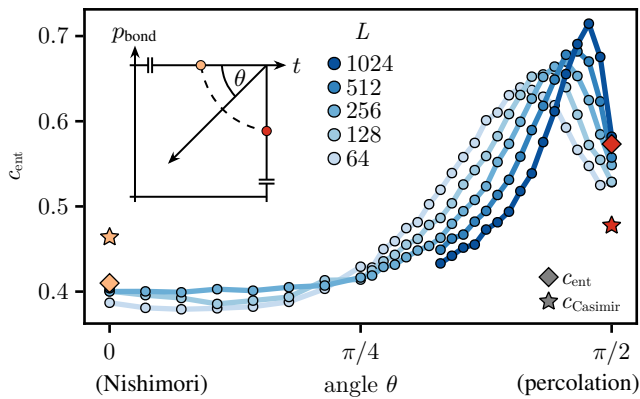}
   \caption{\textbf{Effective central charge on the critical line.} 
   Inset:
      The angle $\theta$ describes the position on the critical line. The Nishimori point corresponds to $\theta = 0$, 
  whereas the percolation point corresponds to $\theta = \pi/2$. 
      The effective central charge governing the average entanglement entropy at Nishimori criticality is numerically computed to be
      $c_\text{ent} = 0.41956(3)$
  (orange diamond), see Fig.~\ref{fig:fig5multifractal}.
      For reference we also show the effective central charge determined by the Casimir energy at the Nishimori point 
      $c_\text{Casimir} = 0.464(4)$~\cite{Pujol2001} (orange star).
      At the percolation critical point, our numerically determined effective central charge agrees with the
      analytically known $c_\text{ent} = \frac{3\sqrt{3}\ln(2)}{2\pi}=0.573\ldots$~\cite{Lang_2020} (red diamond),
      distinct from the Casimir energy estimate $c_{\rm Casimir} = \frac{5 \sqrt{3}}{4 \pi} \ln 2 = 0.478\ldots$ (red star).     
   }
   \label{fig:eff-central-charge}
\end{figure}
%%%%%%%%%%%%%%%%%%%%%%%%%%%%%%%%%%%%%%%%

%%%%%%%%%%%%%%%%%%%%%%%%%%%%%%%%%%%%%%%%%%%%%%%%%%%%%%%%%%%%%%%%%%%%%%%%%%%%
% Entanglement entropy of 1D quantum state
%%%%%%%%%%%%%%%%%%%%%%%%%%%%%%%%%%%%%%%%%%%%%%%%%%%%%%%%%%%%%%%%%%%%%%%%%%%%
\subsection*{Entanglement entropy of 1D quantum state}
In discussing the global phase diagram of Fig.~\ref{fig:schematic}(a) our interest lies on the critical theories describing the transition along the line of 
thresholds, connecting the Nishimori point in the limit of $p_{\text{meas}}=1$ where all bonds are being subject to measurements, and the percolation transition
corresponding to
the Clifford limit of projective (i.e. ``strong'') measurements $t=\pi/4$. 
Both critical points, the Nishimori point at $p_{\rm meas}=1$ and the percolation point at $p_{\rm meas}=1/2$ are monitored quantum systems known to be described by non-unitary CFTs. The non-unitarity of the underlying CFTs is due to the randomness of the measurement outcomes: in general, CFTs describing transitions in monitored systems are known to have central charge $c=0$ -- see {\it e.g.} the recent review~\cite{Potter21review} for a detailed discussion of this point. 
We now discuss 
two universal characteristics of the corresponding non-unitary, logarithmic CFTs~\cite{GURARIE1993, Ludwig05c0log, GURARIE1999, Vasseur_2012, Cardy2013log, Pixley22MIPTCFT, Vasseur24boundarytrsf,  Wang25selfdual}. These are
(i): the so-called ``effective central charge'', which is a universal critical finite-size scaling amplitude quantifying the Casimir effect of the monitored (i.e.\ random) CFT, which we here denote by $c_{\rm Casimir}$, and (ii): the universal coefficient of the logarithm of subsystem size of the von Neumann entanglement entropy, which is (twice) the (typical) scaling dimension of a so-called boundary condition changing ``twist operator'', one located at each end of the entanglement interval~\cite{Hayden16,Vasseur2019,Altman2020weak,Ludwig2020}, which we parametrize 
here by $c_{\rm ent}$. In a monitored (and, more generally, any random) system, the two universal characteristics are in general not equal to each other, i.e.
$c_{\rm Casimir} \not =$
$c_{\rm ent}$.
Often, both are  {\it incorrectly} referred to as  ``effective central charge'', which easily leads to conceptual confusion. Such confusion originates from the fact 
that for  ground states of a completely standard, {\rm non-random}, i.e.\ entirely uniform and translationally invariant  unitary CFT, these two universal characteristics coincide, and are equal to the central charge of the unitary CFT.
However, in the case of monitoring, or for that matter any source of randomesss which by definition breaks translational invariance, 
the central charge of the non-unitary CFT simply 
vanishes~\footnote{In a formulation using replicas, or any equivalent formulation  that permits a description in  terms of field theory, here specifically in terms of conformal field theory, that allows for a definition of a notion of central charge.}, and thus carries no information whatsoever which would be capable of distinguishing different critical points. The universal (non-equal) quantities $c_{\rm Casimir}$ and $c_{\rm ent}$,  on the other hand, are in general non-vanishing in those latter cases, and are powerful universal charactistics distinguishing {\it different} monitored (or, in general, random) non-unitary CFTs.
 
Let us start our discussion with
the quantity $c_{\rm ent}$: First let us briefly review the case of  the ground state of a  standard, translationally invariant CFT. 
In our model, such a unitary CFT description is relevant to a post-selection scenario with uniform $(s=+1)$ 
which realizes a disorder-free (1+1)D critical Ising chain with central charge $1/2$ in lieu of the Nishimori critical point
upon tuning the measurement strength in the limit of $p_{\text{meas}}=1$. 
In this case the von Neumann entanglement entropy is well known~\cite{Calabrese2004,Calabrese2009} to take the form (assuming periodic boundary conditions)
\begin{equation} 
	    S_{vN} = \frac{c}{3} \ln\left [\frac{L}{\pi a}\sin(\frac{\pi l}{L})\right] +\cdots \ ,
	\label{UnitaryCFTEntropy}
\end{equation}
where $c=1/2$ for the post-selected Ising case above, 
the lattice spacing (UV cutoff) $a$ indicating non-universal additive contributions to the entanglement entropy (additive terms indicated by an ellipsis). In the following, we will set $a=1$. A similar expression holds in the case of open boundary conditions where a partitioned interval of size $l$
%\redsout{the entanglement interval of size} %$\ell$ 
sits at the boundary of the system, with a prefactor of the logarithm equal to $c/6$ instead of $c/3$. This expression relies solely on the fact that the entanglement entropy is related to the logarithm of the  two-point function of the twist-operator, which decays as a power of the chord distance $R = \left [\frac{L}{\pi}\sin(\frac{\pi l}{L})\right]$, indicated in 
Eq.~(\ref{UnitaryCFTEntropy}), due to periodic boundary conditions in spatial size $L$.
Now moving on to the case of monitoring of interest to us, the expression for the Born-rule averaged entanglement entropy is of the same form
\begin{equation} 
	\sum_\mathbf{s} P(\mathbf{s}) S_{vN}(\ket{\phi(\mathbf{s})}) = \frac{c_{\text{ent}}}{3} \ln R +\cdots\ ,
	\label{eqAvgEnt}
\end{equation}
where the 1D quantum state $|\phi({\bf s})\rangle$ was defined in Eq.~(\ref{LabelEqDefPhis}).
The reason for this form is that again, the entropy is given by the logarithm of the  two-point function of the twist operator, but now that twist operator correlation function is random
as it depends on the quantum trajectory (it is thus a more complex object), and has to be averaged with the Born rule probability $P(\mathbf{s})$. 
The result turns out~\footnote{For the details of  an explicit derivation of this type of expression see e.g.
Ref.~\onlinecite{Ludwig24stat}.}
to be the right hand side of Eq.~(\ref{eqAvgEnt}).
Here the coefficient $c_{\rm ent}$ determines the {\it average} entanglement entropy of the ensemble of {\it 1D quantum states}. 
That is, for the Born average scenario, we compute the entanglement entropy for each 
set of measurement outcomes ${\bf s}$ yielding
$S_{vN}(\ket{\phi(\mathbf{s})})$,  and average 
it according to the Born probability.

Let us now turn to a discussion of $c_{\rm Casmir}$.
For disordered systems, a quantity directly related to the central charge, called the {\it effective central charge}, is instead given by the universal prefactor governing the Casimir energy~\cite{Ludwig1987,Cardy1998,Zabalo2022}. 
Note that here, the Casimir energy is random depending on the quantum trajectory. The correct quantity to consider turns out to be the {\it averaged} Casimir energy, related to the fact that this is a self-averaging quantity.
In the language of replicas, the effective central charge is given by the derivative of the central charge of the replicated theory 
with respect to the number of replicas
in the replica limit {(which precisely describes~\cite{Ludwig1987} the {\it average} of the Casimir energy in the language of replicas)}: although the central charge becomes zero in the replica limit, its first derivative is in general non-zero [it is also universal since the central charge is universal for any (sufficiently small) number of replicas].
For instance, in the case of percolation, 
the following analytical results are known,
\begin{align}
	c_{\rm ent} = \frac{3 \sqrt{3}}{2 \pi} \ln 2 = 0.573 ,&& c_{\rm Casimir} = \frac{5 \sqrt{3}}{4 \pi} \ln 2 = 0.478 \,,
\end{align}
with $c_{\rm ent}$  
first derived in Ref.~\cite{Saleur08}
(see also Ref.~\onlinecite{Lang_2020,Vasseur24transfer} for a later discussion),
while $c_{\rm Casimir}$ has been 
listed in Refs.~\onlinecite{Cardy1998, Vasseur24transfer,Picco2006}.
Similarly, for the Nishimori critical point (in the absence of bond dilution), those two universal numbers
have been obtained numerically to be
\begin{align}
	c_{\rm ent} = 0.41956(3) \ ,&& c_{\rm Casimir} = 0.464(4) \,,
\end{align}
where the Casimir estimate, $c_{\rm Casimir}$, was obtained
in classical Monte Carlo simulations via scaling 
of the free energy of the classical statistical mechanics model formulation
in Ref.~\onlinecite{Pujol2001}. 
Our high-precision numerical estimate of $c_{\rm ent}$ reported in the paper at hand
is roughly consistent with the one previously obtained by TEBD calculation (albeit only for system sizes up to
$L\sim 300$) 
in Ref.~\cite{Nishino20}. 

Looking at the four estimates above, it is clear that they describe two distinct universality classes. 
But while the Casimir central charge estimates $c_{\rm Casimir}$ for the two critical theories differ only by a few percent,
there is a substantial (some $40\%$) 
difference for the entanglement entropy prefactor $c_{\text{ent}}$ between the two universality classes, 
making the latter the go-to discriminator between the two theories.

%%%%%%%%%%%%%%%%%%%%%%%%%%%%%%%%%%%%%%%%%%%%%%%%%%%%%%%%%%%%%%%%%%%%%%%%%%%%
% RG flow
%%%%%%%%%%%%%%%%%%%%%%%%%%%%%%%%%%%%%%%%%%%%%%%%%%%%%%%%%%%%%%%%%%%%%%%%%%%%
\subsubsection*{RG flow along the critical line}
Let us now shift our attention towards the critical line connecting the percolation and Nishimori critical points in our phase diagram in Fig.~\ref{fig:schematic}(a).
Along this line we have numerically calculated  $c_{\rm ent}$ as shown in Fig.~\ref{fig:eff-central-charge} for systems of varying size with up to $N=1024\times2048$ qubits.
What we qualitatively find is that upon moving away from the Nishimori critical point upon introducing dilute measurements, our numerical  $c_{\rm ent}$ estimate remains constant
at first, indicating that the critical theory remains within the Nishimori universality class. But upon approaching the percolation point, a dramatic shift sets in with $c_{\rm ent}$ 
strongly increasing, overshooting the percolation estimate, which is ultimately approached as one nears the Clifford limit of projective measurements.
This non-monotonous behavior of $c_{\rm ent}$ along the critical line of non-unitary theories should be contrasted to the strict monotonicity dictated by the $c$-theorem
for uniform, translationally invariant unitary CFTs~\cite{ctheorem}, but which 
generally does not apply to the non-unitary case.
Indeed, it has been shown that entanglement does not decrease monotonically under renormalization flows for certain models, see {\it e.g.}~\cite{PhysRevLett.112.106601, Swingle_2014}.
One way to rationalize this non-monotonous behavior in our model is to realize that when moving away from the percolation transition, our system leaves the constrained Hilbert 
space available to
the Clifford circuit
which allows the wavefunction to suddenly populate an extensive number of states in the full Hilbert space.
Nevertheless, the non-Clifford perturbation turns the otherwise maximally entangling gate to a non-maximally entangling gate, which on the microscopic level would only indicate the decrease of the entanglement instead. Therefore the sudden increase of the $c_{\rm ent}$ should arise as a many-body macroscopic phenomenon, which remains to be studied in future work.

%%%%%%%%%%%%%%%%%%%%%%%%%%%%%%%%%%%%%%%%
\begin{figure}[t]
	\centering
	\input{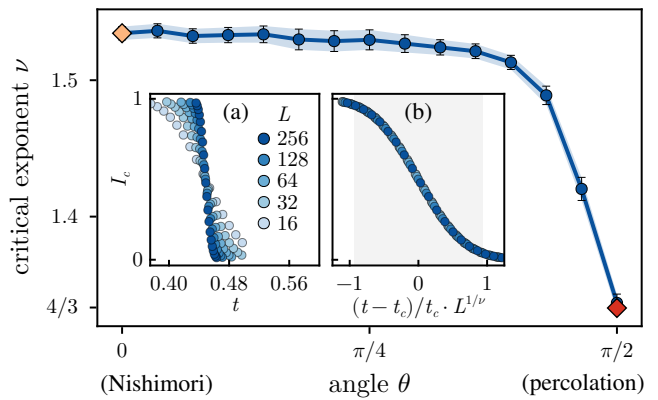}
	\caption{\textbf{Critical exponent $\nu$ on the critical line.} 
		We calculate the critical exponent $\nu$ for different angles $\theta$ on the critical line by finite-size scaling for the coherent information of the surface code (with code distance $L=L_x-1= L_y$ up to $L = 512$ on the Nishimori line and up to $L=256$ elsewhere). 
		At the Nishimori point we obtain a value of $\nu = 1.532(4)$ (orange diamond). 
		At the percolation point we determine $\nu = 1.337(6)$,
		which is consistent with the analytical value~\cite{Stauffer_1979} $\nu = 4/3$ (red diamond).
		Inset (a) shows the raw data for different system sizes and inset (b) shows their data collapse for the Nishimori point ($\theta=0$). 
      These numerical results clearly indicate that the entire line $\theta<\pi/2$ flows toward Nishimori criticality, in agreement with our theory prediction.
	}
	\label{fig:nu_vs_theta}
\end{figure}
%%%%%%%%%%%%%%%%%%%%%%%%%%%%%%%%%%%%%%%%

Beyond the non-monotonicity, the qualitative behavior of $c_{\rm ent}$ along the critical line allows for the identification of a renormalization group (RG) flow of the critical theory. 
With all the action taking place in proximity of the percolation transition, this is where the RG flow happens -- that is, upon moving away from the percolation critical point the system flows to Nishimori universality. 
This direction of the flow can be established analytically by demonstrating the irrelevance of random bond dilution at the Nishimori critical point, 
using the known numerical results for the correlation length exponents (and a Harris-criterion type argument):
Dilution couples to the thermal/energy operators at the Nishimori point, 
		which have scaling dimensions \cite{Hasenbusch2009} $x_1 = 1.345$ and $x_2 = 1.75$ respectively.
		The scaling dimension of the operator coupling to the strength of dilution is twice as large deduced from the replica trick, i.e.
		$x_{\rm dilution} = 2.69$, which is larger than 2 and thus irrelevant (for the two dimensional scenario we are looking at).
		This is the ``Harris-type criterion''~\cite{Harris1974,Chayes1986}
		stating irrelevance if the dimension of the energy operator is larger than one 
		(or the correlation length exponent associated with it $\nu >1$) -- in two dimensions. 
      We review this argument in Appendix.~\ref{NishimoriStabilityAppendix} using a replica-based approach, but we also note that this result was proved rigorously in Ref.~\onlinecite{Chayes1986}.
        
We also note that previous high-precision calculations on the classical diluted RBIM \cite{Picco2006} have studied the behavior of the $c_{\rm Casimir}$ and through very careful simulations (tuning three parameters in the classical model instead of two parameters in our quantum model) have similarly concluded that percolation must be unstable to a flow to Nishimori criticality.

Moving away from the entanglement diagnostic, let us also analyze the correlation length exponent $\nu$ which corroborates the RG flow as shown in  Fig.~\ref{fig:nu_vs_theta}. 
We compute this exponent via a coherent information estimate in the surface code protocol~\cite{teleportcode}, an observable that shows a finite-size independent crossing point
(i.e.\ a zero scaling dimension) at the transition and whose data collapse therefore allows for high-precision estimates of the correlation length exponent.

%%%%%%%%%%%%%%%%%%%%%%%%%%%%%%%%%%%%%%%%%%%%%%%%%%%%%%%%%%%%%%%%%%%%%%%%%%%%
% Multifractality
%%%%%%%%%%%%%%%%%%%%%%%%%%%%%%%%%%%%%%%%%%%%%%%%%%%%%%%%%%%%%%%%%%%%%%%%%%%%
\subsubsection*{Multifractality}

To further analyze the universal properties of the percolation and Nishimori critical points, we study  their ``multifractal'' spectra of scaling dimensions. Since measurement outcomes are random, the monitored systems studied in this paper are intrinsically {\it disordered}. The corresponding critical points and conformal field theories are much richer than their clean, unitary counterparts, as one can distinguish, for example, mean and typical correlation functions. 
More generally, different moments of correlations functions
scale with distinct non-trivial exponents~\cite{LUDWIG1990639}, a phenomenon known as multifractality 
(sometimes also referred to as `multiscaling')
in the literature of disordered critical points. 
This leads to a {\it continuous} spectrum of universal critical exponents (scaling dimensions). We will focus on the scaling behavior of
the boundary condition changing  (boundary) twist operators~\cite{Vasseur2019,Jian2020} whose typical scaling dimensions give rise to the universal prefactors in the R\'enyi entropies, which we referred to as $c_{\rm ent}$ for the von Neumann entropy and which we 
denote by $c_{\rm ent}^{(n)}$ for the $n$th R\'enyi entropy. The  multifractal scaling behavior quantifies the universal scaling behavior of the {\it statistical fluctuations} of these entropies. There is an infinite set of universal quantities for each of 
the $n$th R\'enyi entropies characterizing these fluctuations. Here we follow the  discussion in Ref.~\cite{Ludwig24stat} where this is spelled out in detail for the $n=2$nd R\'enyi entropy  (the discussion being analogous for general $n$), see Appendix~\ref{MultifractalAppendix} for a self-contained summary. In the following we discuss and numerically compute these statistical fluctuations for the von Neumann entropy.
In the previous section we have discussed the Born average of the boundary von Neumann entanglement entropies, which reveals only the first moment of the random variable $\{S_{vN}(\mathbf{s})\}$ (here we abbreviate 
$\ket{\phi({\bf s})}$ by $\ket{\phi}$
inside the argument). It turns out that higher cumulants
(with $\kappa_m$ denoting the $m^{\rm th}$ cumulant)
of this quantity also scale logarithmically at criticality, with universal prefactors,
\begin{equation}
	\begin{split}
	\kappa_2&= \mu_2-\mu_1^2 = -2 x^{(2)} \ln R \ ,\\
	\kappa_3 &= \mu_3 - 3\mu_2\mu_1 + 2\mu_1^3 = 2x^{(3)} \ln R \ ,
	\end{split}
	\label{LabelEqCumulants}
\end{equation}
and more generally
\begin{equation}
	 \kappa_m =(-1)^{m-1} 2 x^{(m)} \ln R \,,
	 \label{eq:kappa_m}
\end{equation} 
with $\mu_m=\sum_\mathbf{s} P(\mathbf{s}) 
[S_{vN} (\mathbf{s})]^m$ being the $m$-th moment average 
and $R$ the chord length. 
Here $x^{(m)}$ adopts the notation from Ref.~\cite{Ludwig24stat}, which expresses the 
scaling of the cumulants of the entanglement entropy (a random variable) with 
$\ln R$, for different cumulant orders $m$, and $x^{(1)}=c_{\text{ent}}/6$ in our notation. 
%
%%%%%%%%%%%%%%%%%%%%%%%%%%%%%%%%
\begin{figure}[t]
	\centering
   \input{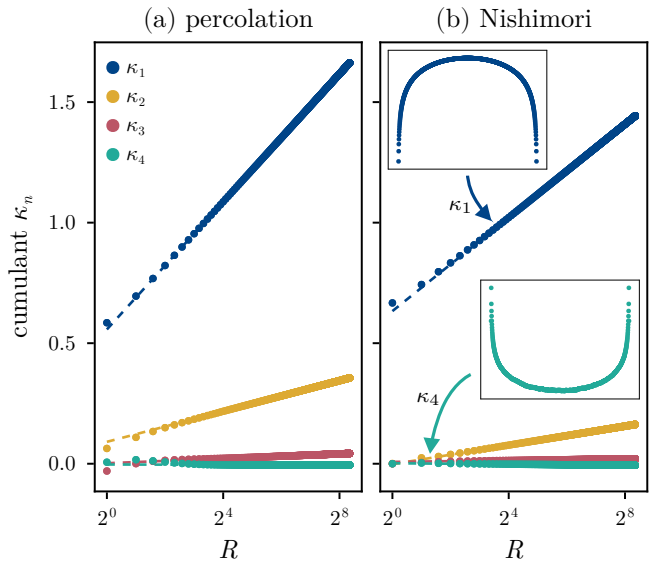} 
	\caption{\textbf{Multifractality} of percolation and Nishimori criticality, which are computed for the representative fixed points to increase numerical accuracy. 
		Shown are the first four cumulants, $\kappa_1,  \kappa_2, \kappa_3$ and $\kappa_4$  of the von Neumann entropy 
		as a function of chord distance. Note that the fourth-order cumulant $\kappa_4$ shows a sign-flip with an inverted `entanglement arc' 
		[cf.\ Eq.~\eqref{UnitaryCFTEntropy}] as shown in the inset.
	}
	\label{fig:fig5multifractal}
\end{figure}
%%%%%%%%%%%%%%%%%%%%%%%%%%%%%%%%

%%%%%%%%%%%%%%%%%%%%%%%%%%%%%%%%
\begin{figure}[t]
	\centering
   \includegraphics[width=\columnwidth]{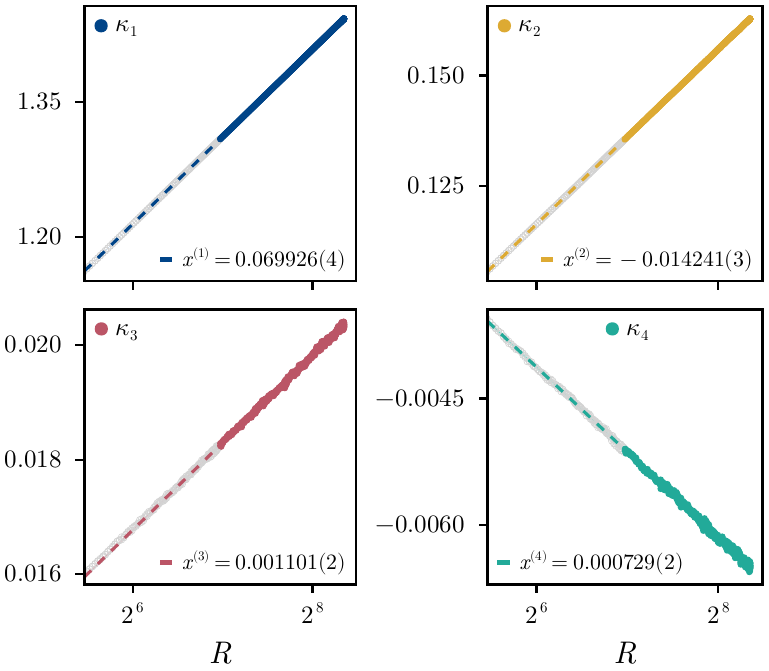}
	\caption{\textbf{Cumulant fits for Nishimori universality.}
		Shown is a subset of the data of Fig.~\ref{fig:fig5multifractal}(b), with the color-highlighted data used in the fits indicated by the dashed lines
		over the remaining data (gray). 
		The data is fitted using Eq.~\eqref{LabelEqCumulants} to obtain the numerical estimates indicated in the figure 
		(and listed in Table \ref{tab:criticalpoints}) 
		To minimize boundary effects we fit only the data in an intermediate window of large cut length $L_x/8 < l< 7L_x/8$, 
		see Appendix \ref{app:cumulants} for supplemental data and a detailed discussion. 
   }
	\label{fig:fig6multifractal_nishimori}
\end{figure}
%%%%%%%%%%%%%%%%%%%%%%%%%%%%%%%%

%
As already mentioned above, different
from Ref.~\cite{Ludwig24stat} here we mainly consider the von Neumann entropy  in the main text, 
while leaving a discussion of the higher-order Rényi entropies for the Appendix. 
In Fig.~\ref{fig:fig5multifractal}, we show the  
scaling of the cumulants
with chord length which yields a spectrum of universal numbers $x^{(m)}$ that are distinct between Nishimori and percolation criticality as summarized in Table \ref{tab:criticalpoints}. Remarkably, contrary to what was previously assumed, even percolation exhibits a non-trivial multifractal spectrum with non-trivial values
for the set of universal numbers $x^{(m)}$. The multifractal behavior of entanglement in the percolation theory can be understood as follows: entanglement in this theory is related to the number of percolation clusters connecting the entanglement interval to its complement. 
This quantity has a non-trivial statistics, which was computed exactly in Ref.~\cite{Saleur08} using Coulomb gas methods. 
The results can be conveniently expressed in terms of the $k$-th moment average
of the $n$-th order generalized purity
(defined in 
Eq.~(\ref{LabelEqMomentsGeneralizedPurity})
 below, the standard purity corresponding to $n=2$)
 in the Born ensemble
\begin{equation}
	\label{LabelEqMomentsGeneralizedPurity}
	\overline{[\text{tr} (\rho(\mathbf{s}))^n]^k} 
	= 
	\overline{e^{-k (n-1) S_n (\mathbf{s})}} 
	\propto 
	R^{-2 X_{n,k}} \,,
\end{equation}
where the overbar denotes the average with respect to the Born rule probability distribution for the measurement outcomes ``$(\bf s)$''. 
Eq.~(\ref{LabelEqMomentsGeneralizedPurity}) describes the $k$-th moment of a boundary two-point function of boundary-condition
changing (bcc) ``twist'' operators~\cite{Vasseur2019,Jian2020,Ludwig24stat} separated by the chord-length $R$.
%
%%%%%%%%%%%%%%%%%%%%%%%%%%%%%%%%
\begin{table*}[t]
\centering
\begin{tabular}[t]{c | ll | llll}
   \toprule %%%%%%%%%%%%%%%%%%%%%%%%%%%%%%%%
   $\quad$ criticality $\quad$ &
   \multicolumn{2}{c|}{$\quad\quad\quad\quad$ \bf percolation $\quad\quad\quad\quad$} &
   \multicolumn{4}{c}{$\quad\quad\quad\quad\quad\quad$ \bf Nishimori $\quad\quad\quad\quad\quad\quad$} \\
   \midrule %%%%%%%%%%%%%%%%%%%%%%%%%%%%%%%%
   $\nu$                  & \multicolumn{2}{c|}{4/3}           & \multicolumn{4}{c}{1.532(4)}     \\
   \midrule %%%%%%%%%%%%%%%%%%%%%%%%%%%%%%%%
   $c_{\rm ent}^{\rm vN}$ & \multicolumn{2}{c|}{$0.573\ldots$} & \multicolumn{4}{c}{$0.41956(3)$} \\
   $c_{\rm Casimir}$      & \multicolumn{2}{c|}{$0.478\ldots$} & \multicolumn{4}{c}{$0.464(4)$}   \\
   \midrule %%%%%%%%%%%%%%%%%%%%%%%%%%%%%%%%
   cumulants              & \mc{$\quad$analytical$\quad$}	& \multicolumn{1}{c|}{numerical}	& \mc{von Neumann}                 & \mc{Rényi-2}         & \mc{Rényi-3}          & \mc{Rényi-$\infty$}      \\
   $x^{(1)}$              & $\quad\pmin0.09554$             & $\pmin0.09556(1)$    			   & $\pmin0.069926(4)$ 	       & $\pmin0.054841(4)$ 	& $\pmin0.049519(3)$ 	& $\pmin0.038573(3)$       \\
   $x^{(2)}$              & $\quad-0.02291$                 & $-0.02289(1)$        			   & $-0.014241(3)$     	       & $-0.012317(2)$       & $-0.011287(2)$        & $-0.007834(1)$           \\
   $x^{(3)}$              & $\quad\pmin 0.00349$            & $\pmin0.00348(1)$    			   & $\pmin0.001101(2))$ 	       & $\pmin0.001433(2)$ 	& $\pmin0.001517(1)$ 	& $\pmin0.001231(1)$       \\
   $x^{(4)}$              & $\quad\pmin0.00023$             & $\pmin0.00021(2)^*$     			& $\pmin0.000729(2)$ 	       & $\pmin0.000644(1)$ 	& $\pmin0.000557(1)$ 	& $\pmin0.000200(1)$       \\
   $x^{(5)}$              & $\quad -0.00026$                & \multicolumn{1}{c|}{---}       & $\pmin0.000050(2)^{\dagger}$ & $-0.000128(1)$	      & $-0.000209(1)$     	& $-0.000189(1)$           \\
   \bottomrule %%%%%%%%%%%%%%%%%%%%%%%%%%%%%%%%
\end{tabular}
\caption{	{\bf Characterization of critical points.}
	For percolation, we list analytical results for the critical exponents, effective and entanglement central charge estimates, 
	along with the five leading cumulants for the $n=1$ von Neumann entanglement entropy, as defined in Eq.~\eqref{eq:kappa_m}. 
	We also provide numerical estimate for these cumulants, extracted from fits like the ones shown in Fig.~\ref{fig:fig5multifractal}(a).
	The asterisk ($^*$) for the fourth-order cumulant indicates that we needed to ``cherry-pick" the fitting interval in order to reproduce 
	a value close to the analytical estimate (see Fig.~\ref{fig:multifractality_percolation_nishimori_periodic_boundary} of the Appendix).
	For Nishimori criticality, no analytical results are known and we list our numerical estimates. 
	Note that for the cumulant estimates we are able to obtain high-precision results for the leading four cumulants from the fits shown in
	Fig.~\ref{fig:fig6multifractal_nishimori} and even a rough estimate for the fifth-order one 
	(see 	Fig.~\ref{fig:multifractality_percolation_nishimori_periodic_boundary}(b) of the Appendix for the underlying fit). 
	These Nishimori cumulants we have calculated not only from the von Neumann entanglement entropy but also the second, third
	and $n=\infty$ order R\'enyi entropies as indicated.
	The dagger ($\dagger$) for the fifth-order von Neumann cumulant of Nishimori criticality indicates that it is likely unreliable
	given that there is a ``flip" in the entanglement arc that might arise from boundary or finite-size effects, 
	see Fig.~\ref{fig:multifractality_percolation_nishimori_periodic_boundary} in the Appendix.
}
\label{tab:criticalpoints}
\end{table*}
%%%%%%%%%%%%%%%%%%%%%%%%%%%%%%%%
For percolation (which can be formulated as a particular Clifford circuit \cite{Nahum20freefermion,Lang_2020,Vasseur24transfer}),
this quantity does not depend on the R\'enyi index $n$, and the scaling dimension $X_{n,k} = X_k$ is the generating function $X_k = \sum_m \frac{x^{(m)}}{m!} k^m $ of the universal coefficients appearing in Eq.~\eqref{LabelEqCumulants}~\cite{Ludwig24stat}. 
In this case we may choose $n=2$, and  Eq.~(\ref{LabelEqMomentsGeneralizedPurity}) describes the $k$-th moment of the 
(standard) purity.
Using the results of Ref.~\cite{Saleur08}, we find that for percolation, we have the
exact multi\-fractal spectrum of critical exponents
\begin{equation}
	X_k =  \frac{3 \arccos^2 (\sqrt{3}/2^{1+k/2})}{2\pi^2} -\frac{1}{24} \,.
	\label{eq:percolation_exponents}
\end{equation}
with
\begin{eqnarray}
	x^{(1)} & = & c_{\text{ent}}/6 = \frac{\sqrt{3}}{4 \pi} \ln 2   \simeq 0.09554 \, , \nonumber \\
	x^{(2)} & = & - 0.022914  \, , \nonumber \\
	x^{(3)} & = & \phantom{-} 0.0034922 \, ,  \nonumber \\
	x^{(4)} & = & \phantom{-}0.0002325 \, ,  \nonumber \\
	x^{(5)} & = & -0.0002561 \, .
\end{eqnarray}
Note that with our conventions [Eq.~\eqref{eq:kappa_m}], $x^{(2)}$ is negative by definition since the variance must be positive. 
Further note that the positiveness of $x^{(4)}$ predicts that the 4th cumulant $\kappa_4 = -2x^{(4)}\ln R$ 
scales, following Eq.~\eqref{UnitaryCFTEntropy}, as an inverted `entanglement arc' (see the inset of Fig.~\ref{fig:fig5multifractal} for an example).
Extracting numerically the asymptotic behavior of higher cumulants is extremely challenging as it requires multiple cancellations between 
powers $(\ln R)^k$ where the chord length $R$ is large in the limit of interest in Eq.~\eqref{LabelEqCumulants}, 
and we need to extract the coefficient of the lowest, first-order power of this large quantity $\ln R$ [see Eq.~\eqref{LabelEqCumulants}].
Nevertheless, we find that for our numerical results for critical percolation
[Fig.~\ref{fig:fig5multifractal}(a)] the first two significant digits  
generally agree well with the exact results, 
see also the side-by-side comparison of Table~\ref{tab:criticalpoints} above. For the percolation problem, fitting the fourth-order cumulant remains a 
difficult endeavor and, to some extent, we have `cherry-picked' the fitting range to match the analytical expectation 
(see Fig.~\ref{fig:multifractality_percolation_nishimori_periodic_boundary} in the Appendix for further details).

Surprisingly, it turns out that the numerical estimates for Nishimori criticality, shown in Fig.~\ref{fig:fig5multifractal}(b) 
and also summarized in Table~\ref{tab:criticalpoints}, converge significantly faster -- intuitively, this makes sense since the Clifford/percolation data involves averaging {\it discrete} quantities, and thus requires more samples. Our analysis of the von Neumann entropy
yields the following values 
\begin{eqnarray}
	x^{(1)} & = & \phantom{-} 0.069926(4) \, , \nonumber \\
	x^{(2)} & = & - 0.014241(3) \, , \nonumber \\
	x^{(3)} & = & \phantom{-} 0.001101(2) \, ,  \nonumber \\
	x^{(4)} & = & \phantom{-} 0.000729(2) \, ,  
\end{eqnarray}
which are also listed in Table \ref{tab:criticalpoints} in the column denoted von Neumann and compared to results from the second, third, and $n=\infty$ 
R\'enyi entropies as indicated in the column headers. \\

To summarize these results, it is quite notable that the two monitored (non-unitary) critical points studied here, percolation and Nishimori criticality, 
both contain an {\it infinite} set of scaling dimensions that constitute a {\it continuous} spectrum
in contrast to conventional (unitary) critical points. 
This richness of critical exponents originates from the random nature of monitored criticality and reflects their multifractality.
For percolation, we have been able to compute the entire (non-quadratic) multifractal spectrum exactly, while for Nishimori criticality our numerical simulations allowed to access the leading five cumulants.

\vskip 1cm

%%%%%%%%%%%%%%%%%%%%%%%%%%%%%%%%%%%%%%%%%%%%%%%%%%%%%%%%%%%%%%%%%%%%%%%%%%%%
% Discussion 
%%%%%%%%%%%%%%%%%%%%%%%%%%%%%%%%%%%%%%%%%%%%%%%%%%%%%%%%%%%%%%%%%%%%%%%%%%%%

\subsection*{Discussion}

It is conceptually important  to note the following two equivalent roles that percolation plays in our work: we have so-far employed percolation as a means to simply describe the {\it geometrically} and purely classical random dilution of bonds of the 2D lattice. On this diluted lattice the RBIM quantum mechanics (viewed as a shallow quantum circuit on a 2D spatial lattice, in an 
Rokhsar-Kivelson (RK) formulation as mentioned above) is located. On the other hand, percolation itself can be viewed as a deep (1+1)D random Clifford circuit of projective measurements on the 1D lattice~\cite{Nahum20freefermion,Lang_2020,Hsieh2021measure, Barkeshli2021measure, Vasseur24transfer} whose spacetime is the above 2D lattice. The 2D quantum states on the critical, geometric percolation cluster, i.e. at $p_{\rm meas}=1/2$ and $t=\pi/4$ in our phase diagram in Fig.~\ref{fig:schematic}, can be exactly mapped to such a deep (1+1)D Clifford circuit with random projective measurements, which represents the geometrical percolation problem in spacetime in a quantum mechanical (1+1)D deep circuit language. 
The percolation criticality is well known to have a flat entanglement spectrum
because it can be thought of as a deep Clifford circuit~\cite{Nahum20freefermion,Lang_2020,Vasseur24transfer} and thus 
all R\'enyi entropies are trivially identical in every quantum trajectory.
Nevertheless, here we show analytically that the statistical ensemble of the entanglement entropies over different quantum trajectories exhibits
{\it multifractality}, which is further evidenced by our numerical computation 
of the cumulants of these entanglement entropies up to 3rd, 4th, and 5th cumulants, that are elusive in the literature.  

Moving away from the point of percolation criticality $(p_{\rm meas}=1/2$, $t=\pi/4)$
in our phase diagram  along the critical RG crossover line, is a concrete way of ``turning on'' the non-Clifford perturbations at the percolation critical point of the (1+1)D deep Clifford circuits. These arise from the additional quantum mechanics of the Nishimori measurements, which can be induced by weakening the projective measurements~\cite{NishimoriCat}, 
or by injecting coherent error prior to measurements~\cite{Chen25nishimori, Beri23toriccodecoherenterror, Beri24coherenterror, Gullans24magic, Ippolitti24coherdesign, Beri241022436} in the Clifford circuit. 

%%%%%%%%%%%%%%%%%%%%%%%%%%%%%%%%%%%%%%%%%%%%%%%%%%%%%%%%%%%%%%%%%%%%%%%%%%%%
\begin{figure}[b]
\includegraphics[width=\columnwidth]{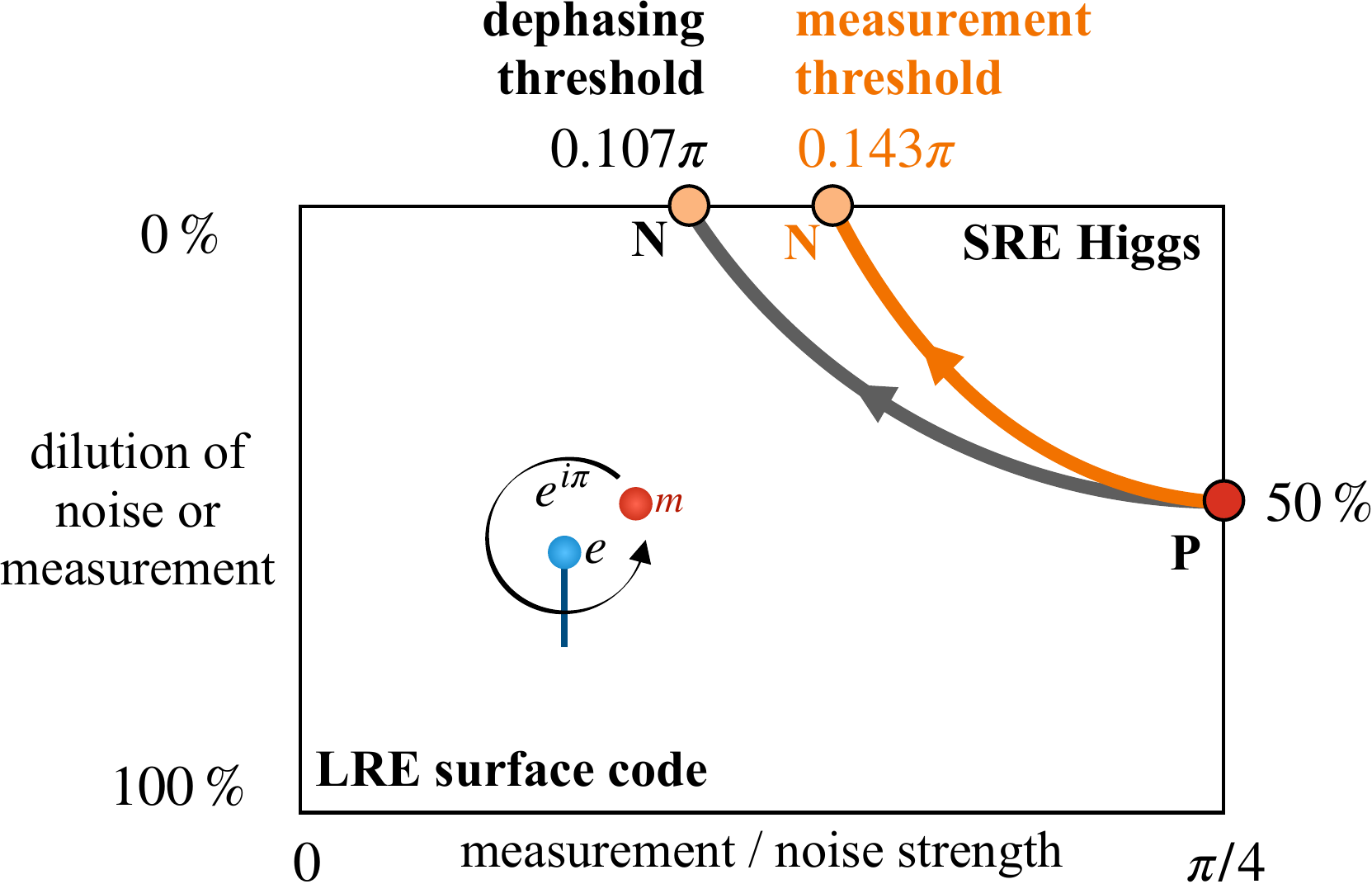}
\caption{{\bf Schematic phase diagram of diluted noise channel in comparison with diluted measurement} of the topological toric code or surface code. Both exhibit a critical line from percolation to the Nishimori points, albeit at different threshold values. Here the horizontal axis is the measurement strength $t$ for the measurement protocol. For noisy states, the horizontal axis translates the noise probability to $t$ by $p_{\it noise}=\sin^2(t)$. The vertical axis denotes the dilution of the noise or measurement, such that $0\%$ refers to uniform noise or measurement, and $100\%$ means no noise or measurement occurs. }
\label{fig:noisyphasediagram}
\end{figure}
%%%%%%%%%%%%%%%%%%%%%%%%%%%%%%%%%%%%%%%%%%%%%%%%%%%%%%%%%%%%%%%%%%%%%%%%%%%%

More generally, the measurement-induced phase transitions we reveal here can also be viewed~\cite{Wang25selfdual}
as a mixed state transition for the classical-quantum mixed state $\rho = \sum_\mathbf{s}  \ketbra{\mathbf{s}} \otimes \ketbra{\psi(\mathbf{s})}$, where $\ketbra{\mathbf{s}}$ denotes the state of
ancilla qubits that record the measurement  outcomes,
and we take $\ket{\psi(\mathbf{s})}$ to be the toric code state subjected to diluted weak measurements, as written in 
Eq.~(\ref{eq:psi}). 
It can be shown 
that the {\it same} universality class describes the mixed state subjected to a diluted dephasing noise channel instead of quantum measurement, see Fig.~\ref{fig:noisyphasediagram} for a schematic and Appendix~\ref{sec:appendixA} for details. Here we outline the basic idea as follows:
When the measurement 
outcomes are erased (i.e.\ traced over), the observer acts as a bath, and the ensemble of the post-measurement states is no longer an ensemble of pure states, but simply gives rise to a mixed state.
Effectively, the toric code is subjected to a diluted dephasing channel~\cite{teleportcode}  with noise probability $\sin^2(t)$; see Appendix~\ref{sec:appendixA} for a detailed discussion.
In the non-dilute limit $p_{\rm meas}=1$ with uniform dephasing noise (top line of our phase diagram Fig.~\ref{fig:schematic}), the noisy toric code exhibits a Nishimori transition~\cite{Preskill2002}, 
albeit at a smaller threshold~\footnote{Note that throwing away the measurement record is a channel that is generally expected not to increase the threshold.} $t_c\approx0.107\pi$~\cite{teleportcode} -- see the schematic illustration of  Fig.~\ref{fig:noisyphasediagram}. 
In the maximally noisy limit $t=\pi/4$ with noise hitting only a fraction of the qubits, the state again undergoes the same percolation transition at $p_{\rm meas}=1/2$. Thus, for the noisy state after erasure (tracing out) of the measurement record $\mathbf{s}$, the phase boundary between toric code and dephased toric code extends from the same percolation critical point to a Nishimori critical point (at $t_c\approx 0.107\pi$), which is described by the same RG flow as revealed in this manuscript. \\

%%%%%%%%%%%%%%%%%%%%%%%%%%%%%%%%%%%%%%%%%%%%%%%%%%%%%%%%%%%%%%%%%%%%%%%%%%%%
% Data availability
%%%%%%%%%%%%%%%%%%%%%%%%%%%%%%%%%%%%%%%%%%%%%%%%%%%%%%%%%%%%%%%%%%%%%%%%%%%%

{\it Data availability}.-- 
The numerical data shown in the figures and the data for sweeping the phase diagram is available on Zenodo~\cite{zenodo_percolation}.\\

%%%%%%%%%%%%%%%%%%%%%%%%%%%%%%%%%%%%%%%%%%%%%%%%%%%%%%%%%%%%%%%%%%%%%%%%%%%%
% Acknowledgments
%%%%%%%%%%%%%%%%%%%%%%%%%%%%%%%%%%%%%%%%%%%%%%%%%%%%%%%%%%%%%%%%%%%%%%%%%%%%

\begin{acknowledgments}

The Cologne group gratefully acknowledge partial funding from the Deutsche Forschungsgemeinschaft (DFG, German Research Foundation)
under Germany's Excellence Strategy -- Cluster of Excellence Matter and Light for Quantum Computing (ML4Q) EXC 2004/1 -- 390534769 
and within the CRC network TR 183 (Project Grant No.~277101999) as part of subproject B01.
GYZ acknowledge the support of Start-up Fund of HKUST(GZ) (No. G0101000221).
The numerical simulations were performed on the JUWELS cluster at the Forschungszentrum Juelich.

\end{acknowledgments}

%%%%%%%%%%%%%%%%%%%%%%%%%%%%%%%%%%%%%%%%%%%%%%%%%%%%%%%%%%%%%%%%%%%%%%%%%%%%
% Bibliography
%%%%%%%%%%%%%%%%%%%%%%%%%%%%%%%%%%%%%%%%%%%%%%%%%%%%%%%%%%%%%%%%%%%%%%%%%%%%

\bibliography{measurements}

%%%%%%%%%%%%%%%%%%%%%%%%%%%%%%%%%%%%%%%%%%%%%%%%%%%%%%%%%%%%%%%%%%%%%%%%%%%%
% Appendix
%%%%%%%%%%%%%%%%%%%%%%%%%%%%%%%%%%%%%%%%%%%%%%%%%%%%%%%%%%%%%%%%%%%%%%%%%%%%
\appendix
\vskip 30mm

%%%%%%%%%%%%%%%%%%%%%%%%%%%%%%%%%%%%%%%%%%%%%%%%%%%%%%%%%%%%%%%%%%%%%%%%%%%%
\begin{center}
{\large Supplementary notes}
\end{center}
%%%%%%%%%%%%%%%%%%%%%%%%%%%%%%%%%%%%%%%%%%%%%%%%%%%%%%%%%%%%%%%%%%%%%%%%%%%%

\section{Analytical supplements}
%%%%%%%%%%%%%%%%%%%%%%%%%%%%%%%%%%%%%%%%%%%%%%%%%%%%%%%%%%%%%%%%%%%%%%%%%%%%
\label{sec:appendixA}

%%%%%%%%%%%%%%%%%%%%%%%%%%%%%%%%%%%%%%%%%%%%%%%%%%%%%%%%%%%%%%%%%%%%%%%%%%%%
\subsection{Gauge transformation}
\label{sec:gaugetransform}
%%%%%%%%%%%%%%%%%%%%%%%%%%%%%%%%%%%%%%%%%%%%%%%%%%%%%%%%%%%%%%%%%%%%%%%%%%%%

\begin{figure}[h!] 
   \centering
   \includegraphics[width=\columnwidth]{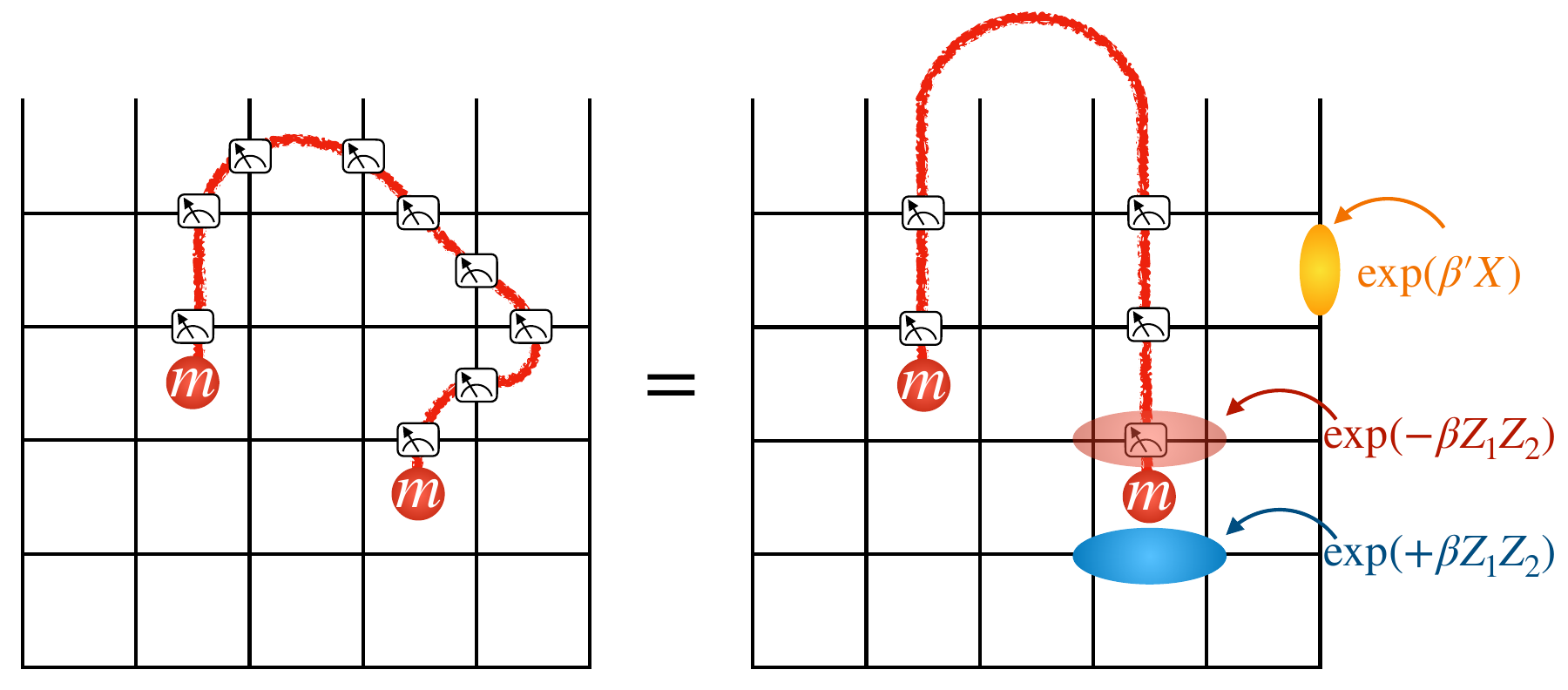} 
   \caption{{\bf Gauge transformation}. We denote the negative measurement outcomes as a string (in red) passing by, along the dual lattice, whose end points define the $m$ vortices. The gauge transformation can fluctuate the shape of the string, without altering the configuration of the $m$ vortices. It relates two distinct measurement outcomes, with identical probability / partition function. The boundary 1+1D states are related by local Pauli $X$ operators and thus share the same entanglement. As a result, we can always turn the string to be straight, such that the $X$ evolution gate is fixed, while the randomness occurs only to the $ZZ$ evolution gates. }
   \label{fig:gaugetransform}
\end{figure}

Here we show the gauge symmetry of the 2D tensor network allows us to relate different measurement outcomes $\mathbf{s}\to \mathbf{s}'$ with identical probability / partition function, see Fig.~\ref{fig:gaugetransform}. In this way, we can fix a temporal gauge and restrict ourselves to the measurement outcomes that have straight vortex strings stretching across the spatial links, such that the corresponding (1+1)D deep circuit has a fixed $X$ evolution gate, while the randomness occurs only to the $ZZ$ evolution gates. Thus the (1+1)D deep circuit requires a post-selection of the $X$ measurement outcomes, or by coupling the system to a bath with dissipation that leads to the imaginary time $X$ evolution, with mid-circuit measurement of the $ZZ$ operators. 
Note that our 2D quantum protocol does not need any (uniform) post-selection. 

%%%%%%%%%%%%%%%%%%%%%%%%%%%%%%%%%%%%%%%%%%%%%%%%%%%%%%%%%%%%%%%%%%%%%%%%%%%%
\subsection{Mixed state from measurement to noise channel}
\label{sec:mixedstate}
%%%%%%%%%%%%%%%%%%%%%%%%%%%%%%%%%%%%%%%%%%%%%%%%%%%%%%%%%%%%%%%%%%%%%%%%%%%%

Here we provide more details on the last paragraph of the discussion section in main text by giving a more detailed derivation of the mixed state
that arises after the measurement record is erased. 
First recall that the toric code under diluted weak measurement can be viewed~\cite{Wang25selfdual} as a mixed state density matrix, 
by taking into account not only the \redsout{system} qubits of the toric code but also the auxiliary qubits that record the measurement outcome.
This mixed state can be written as
\begin{equation}
   \rho = \sum_\mathbf{s} 
   \ketbra{\mathbf{s}} \otimes \ketbra{\psi(\mathbf{s})} \ ,
   \label{eq:rho}
\end{equation}
where $\ket{\psi(\mathbf{s})}$ is the un-normalized post-measurement wave function
from Eq.~(\ref{eq:psi}) of the main text,
\begin{equation}
\ket{\psi(\mathbf{s})} = e^{\frac{\beta}{2} \sum_{\langle ij\rangle \in \mathcal{C}} s_{ij} Z_{ij} }\ket{\psi_0} \ . \nonumber
\end{equation}
Here, $\ket{\psi_0} =\ket{\psi}_{TC}$ is the initial perfect (unmeasured and undiluted) toric code state prior to the measurement, 
as the ground state of the star stabilizer operator $A_v = \prod_{\langle ij\rangle \mid v\in \partial \langle ij\rangle}Z_{ij}$ for 
any vertex $v$, and the plaquette stabilizer operator $B_p = \prod_{\langle ij\rangle\in \partial p} X_{ij}$ for any plaquette $p$
(where $\partial$ denotes the boundary). 
The post-measurement state $\ket{\psi(\mathbf{s})}$, Eq.~(\ref{eq:psi}), is not an eigenstate of the plaquette operator $B_p$. Its norm yields the Born probability of each measurement outcome,
\begin{equation}
   P(\mathbf{s})=\braket{\psi(\mathbf{s})} \ ,
\end{equation} 
which is related to the partition function of a random bond Ising model at inverse temperature $\beta = \tanh^{-1}\sin(2t)$ as defined in 
Eq.~(\ref{eq:P}). From now on, we write more explicitly
\begin{equation}
   P(\mathbf{s}; \beta)\equiv P(\mathbf{s})
   \label{LabelEqPsbeta}
\end{equation} 
to emphasize the $\beta$ dependence of the partition function. 
Here, the quantity $\mathbf{s}$ plays in the mixed state $\rho$ of  Eq.~(\ref{eq:rho}) a role 
similar to a local ``quantum number" that is conserved and specifies %the 
 a block of the density matrix. 
Note that gauge equivalent configurations share the same probability: $P(\mathbf{s}; \beta) = P(\mathbf{s'}; \beta)$ if $\partial\mathbf{s}=\partial\mathbf{s'} = \mathbf{m}$, where we use $\partial \mathbf{s}$ to denote the end point configuration of the strings defined by $s_{ij}=-1$, as with the toric code convention, see Fig.~\ref{fig:gaugetransform} for an example. More concretely, we use $m_p=1$ to label the existence of an excited $m$ vortex on the plaquette $p$, and $m_p=0$ otherwise, by defining $(-1)^{m_p} = \prod_{\langle ij\rangle\in p} s_{ij}$, the latter of which is usually called the Wilson loop operator in the gauge theory language.  
Thus one might as well label the probability function in Eq.~(\ref{LabelEqPsbeta}) by
$P(\mathbf{m}; \beta)$, bearing in mind that $s$ recides on the bond center while $m$ recides on the plaquette center. 
The bulk von Neumann entropy of the mixed state $\rho$ in Eq.~(\ref{eq:rho})
corresponds~\cite{Wang25selfdual} to the Shannon entropy of the measurement record, 
$-\sum_{\mathbf{s}} P(\mathbf{s};\beta)
\ln P(\mathbf{s};\beta)$,
which is by Eq.~(\ref{eq:P})
exactly the quenched free energy of the RBIM, since $P(\mathbf{s}; \beta)$ can be interpreted as the disordered partition function.
This establishes the map to the statistical model and the Nishimori transition point $t_c\approx 0.143\pi$ (Fig.~\ref{fig:schematic} in the main text).

Now we show how the measurement problem can be related to the noise problem. 
If the measurement record is erased (i.e.\ traced out), the mixed state loses $\mathbf{s}$ as a good ``quantum number'', and is converted into
\begin{equation}
   \begin{split}
   \tilde{\rho} &= {\rm tr}_\mathbf{s}\rho =\sum_\mathbf{s}  \ketbra{\psi(\mathbf{s})} = \prod_{\langle ij\rangle\in \mathcal{C}} \mathcal{N}_{ij} (\ketbra{\psi_0})\\
   \end{split}
   \label{eq:mixedstate}
\end{equation}
where $\mathcal{N}_{ij}$ denotes a standard dephasing noise channel acting on a single bond qubit 
$\langle i,j \rangle$
as $\mathcal{N}_{ij}(\ketbra{\psi_0})=(1-\tilde{p}) \ketbra{\psi_0} + \tilde{p} Z_{ij} \ketbra{\psi_0} Z_{ij}$. 
The effective noise probability $\tilde{p}=\sin^2(t)$ is obtained because on each single bond-qubit
\begin{equation}
   \begin{split}
&\mathcal{N}_{ij} (\ketbra{\psi_0}) = \sum_{s_{ij}=\pm 1} e^{\beta s_{ij} Z_{ij}/2}\ketbra{\psi_0} e^{-\beta s_{ij} Z_{ij}/2} \\
&= \cos^2(t)\ketbra{\psi_0} + \sin^2(t)Z_{ij} \ketbra{\psi_0} Z_{ij} \ .
\label{eq:dephasing}
\end{split}
\end{equation} 
In this way, the weak measurement limit $t=0$ corresponds to the noiseless and the projective measurement limit $t=\pi/4$ corresponds to the maximally noisy limit. 
Since the Kraus operator $Z_{ij}$  (at each bond $\langle i,j \rangle$)
acts on both sides of the density matrix, the $m$ (anyon) vortex of the toric code is a conserved quantity and serves as the local ``quantum number'' that replaces the aforementioned
quantity $\mathbf{s}$. The mixed state 
becomes, after tracing out the measurement record,
a grand canonical ensemble of the $m$ vortices,
\begin{equation}
\tilde{\rho} = \sum_\mathbf{m} P(\mathbf{m}; \tilde{\beta}) \ketbra{\tilde{\psi}(\mathbf{m})}
\end{equation}
where $\ket{\tilde{\psi}(\mathbf{m})}$ now denotes a {\it normalized} random toric code eigenstate with eigenvalue $B_p = (-1)^{m_p}$ and $\mathbf{m}$ labels an arbitrary configuration of the $m$ vortices distributed in space. The probability distribution of each such
$\mathbf{m}$
configuration is given by Ref.~\cite{Preskill2002}:
\begin{equation}
   P(\mathbf{m}; \tilde{\beta}) \propto \sum_{\mathbf{s}\mid \partial \mathbf{s} = \mathbf{m}} \left(\frac{p}{1-p} \right)^{\sum_{ij}\frac{1-s_{ij}}{2}} \propto \sum_{\sigma} e^{\tilde{\beta} \sum_{\langle ij\rangle } s_{ij}\sigma_i\sigma_j}\ .
\end{equation} 
As a partition function, it collects the weights of all possible error chains of Pauli $Z_{ij}$ operators on bonds $\langle i,j \rangle$, supported on the string path that ends at the $m$ vortices, 
which is, strictly along the Nishimori line
$\tilde{p}/(1-\tilde{p}) = e^{-2\tilde{\beta}}$,
dual to the partition function of the random bond Ising model.
To summarize, the noisy mixed state $\tilde{\rho}$ is mapped to the RBIM along  the Nishimori line described by the partition function $P(\mathbf{m}; \tilde{\beta})$, 
with disorder probability $\tilde{p}=\sin^2(t)$ and inverse temperature $\tilde{\beta}=\tanh^{-1}\cos(2t)$. 
\\

For comparison, the measurement induced mixed state~\eqref{eq:rho}, thus containing the measurement record $|\mathbf{s}\rangle \langle\mathbf{s}|,$
is mapped to the RBIM along the Nishimori line described by partition function $P(\mathbf{s})$ with disorder probability $p = \cos^2(t)$ and temperature $\beta = \tanh^{-1}\sin(2t)$ in Ref.~\cite{teleportcode}. 
Note the crucial fact that $\tilde{\beta}$ is the Kramers-Wannier dual of $\beta$: $\tilde{\beta} = -\frac{1}{2}\ln \tanh\beta$. 
Namely, erasing the measurement record effectively induces a high temperature to low temperature duality, reversing the temperature of the effective statistical model
along the Nishimori line. Besides, it also turns the non-stabilizer state $\ket{\psi(\mathbf{s})}$ into a stabilizer state $\ket{\tilde{\psi}(\mathbf{m})}$. 

In the absence of dilution, despite the microscopic difference, both the post-measurement states $\rho$ and the noisy states $\tilde{\rho}$ are described by 
Nishimori criticality. 
In the presence of dilution, only a finite fraction of the qubits, supported on the bonds $\langle ij\rangle \in \mathcal{C}$, are subjected to measurement in $\rho$ and dephasing in $\tilde{\rho}$. Here the volume of the support $|\mathcal{C}|=p_{\rm meas}N_B$ is a fraction of the total number of bonds $N_B$ under a fixed probability $p_{\rm meas}$. Note
the crucial fact that this probability $p_{\rm meas}$ plays a different role %with
than $\tilde{p}$ in the dephasing case, because the dilution is a so-called ``heralded'' or ``flagged'' error, which means the information $\mathcal{C}$ is known. 
In contrast, in Eq.~\eqref{eq:dephasing}, one does not have a ``flag'' when $Z_{ij}$ is applied. 
The global phase diagram for such a scenario is left for future study. Nonetheless, at $t=\pi/4$, the percolation transition at $50\%$ still holds. 
Therefore, in such noise phase diagram, there is still a critical line between the exact Nishimori transition at $t=0.107\pi$ and the exact percolation transition at $50\%$, see Fig.~\ref{fig:noisyphasediagram}. We expect the same RG flow between these two critical points.

%%%%%%%%%%%%%%%%%%%%%%%%%%%%%%%%
\begin{table*}[t]
   \centering
   \begin{tabular}[t]{c | c c | c | c | c}
      \toprule %%%%%%%%%%%%%%%%%%%%%%%%%%%%%%%%
      method                               & \multicolumn{3}{c |}{$\qquad\qquad$ tensor networks $\qquad\qquad$} & $\quad$ Clifford simulations $\quad$ & $\quad$ free-fermion network $\quad$                                             \\
      \midrule %%%%%%%%%%%%%%%%%%%%%%%%%%%%%%%%
      Observable                           & \multicolumn{2}{c |}{$I_c$}                                         & $S_{n}$            & $S_{n}$       
      & $S_{n}$                                \\
      \midrule %%%%%%%%%%%%%%%%%%%%%%%%%%%%%%%%
      largest system size $L_x \times L_y$ & $\quad 257 \times 256 \quad$             & $\quad 513 \times 512 \quad$       & $\quad 1024 \times 2048 \quad$   & $\quad 1024 \times 4096 \quad$ & $\quad 1024 \times 20480^* \quad$ \\
      \# of data points                    & $300$                                                               & $20$                   & $15$                 & $1$                & $1$                   \\
      \# of samples / data point           & $100,000$                                                           & $100,000$              & $50,000$             & $14,400,000$       & $18,000,000$          \\
      \# of total core hours               & $650,000$                                                           & $350,000$              & $150,000$            & $200,000$          & $2,300,000$           \\
      total file size                      & $1.3$ GB                                                             & $15$ MB               & $60$ GB              & $220$ GB           & $2$ TB                \\
      \bottomrule %%%%%%%%%%%%%%%%%%%%%%%%%%%%%%%%
   \end{tabular}
   \caption{	{\bf Computational resources.} 
   To characterize the computational effort for our numerical simulations we provide, for the largest system sizes that we have simulated, 
   the number of data points sweeping the phase diagram, number of samples per data point, the total number of core hours, and typical data storage requirements for
   simulations using the numerical techniques indicated at the top of every column.
   The first column lists the numbers for the tensor network simulations of the coherent information $I_c$ throughout the phase diagram. This data is used to calculate the location of the critical line connecting the percolation point and the Nishimori point, as well as determining the critical exponent $\nu$ along that critical line (see Fig. \ref{fig:nu_vs_theta}). Limited to the Nishimori line, we do the same simulation for larger system sizes (second column). We use the tensor network method to calculate the entanglement entropies $S_{n}$ at different points along the critical line in open boundary conditions (third column). This data is used to determine the effective central charge $c_{\rm ent}$ along the critical line (see Fig. \ref{fig:eff-central-charge}) to see the crossover behavior between percolation and Nishimori criticality. In order to calculate the entanglement entropy in periodic (and open) boundary conditions much more efficiently at the percolation point, we use Clifford simulations (fourth column). This data is used to determine the cumulants shown in Fig. \ref{fig:fig5multifractal}(a) and \ref{fig:multifractality_percolation_nishimori_periodic_boundary}(a). Similarly, we use free-fermion network evolution to calculate the entanglement entropies in periodic boundary conditions at the Nishimori point (fifth column). The asterisk ($^*$) indicates, that we calculate the entanglement entropies multiple times while evolving the 1D quantum state, starting after a thermalization time of $4096$ time steps. The data is used to determine the cumulants for the Nishimori point, shown in Figs. \ref{fig:fig5multifractal}(b), \ref{fig:fig6multifractal_nishimori}, \ref{fig:multifractality_percolation_nishimori_periodic_boundary}(b) and \ref{fig:multifractality_nishimori_Renyi}.}
   \label{tab:hpc}
\end{table*}
%%%%%%%%%%%%%%%%%%%%%%%%%%%%%%%%%%%%%%%%%%%%%%%%%%%%%%%%%%%%%%%%%%%%%%%%%%%%

%%%%%%%%%%%%%%%%%%%%%%%%%%%%%%%%%%%%%%%%%%%%%%%%%%%%%%%%%%%%%%%%%%%%%%%%%%%%
\subsection{Coherent information / domain wall entropy}
%%%%%%%%%%%%%%%%%%%%%%%%%%%%%%%%%%%%%%%%%%%%%%%%%%%%%%%%%%%%%%%%%%%%%%%%%%%%

In technical terms, the coherent information for every post-measurement pure state is given
by the measurement (disorder) average of the domain-wall entropy
\begin{equation}
	I_{\mathbf{s}}=-\frac{1+C_{\mathbf{s}}}{2}\log_2 \frac{1+C_{\mathbf{s}}}{2}-\frac{1-C_{\mathbf{s}}}{2}\log_2 \frac{1-C_{\mathbf{s}}}{2} \ ,
\end{equation}
where $C_{\mathbf{s}}$  
the overlap between the two topologically degenerate states of the surface code conditioned upon a given measurement record $\mathbf{s}$. 
In the dual Ising representation,
$C$ is the correlation between two test spins placed on the left and right boundaries of
the lattice~\cite{teleportcode}. The left (right) test spin interacts with all qubits on the left (right) boundary with the same Ising interaction strength, also subjected to the bond randomness. Consequently, $C$ can be viewed as the expectation value of a ``domain wall" operator (with eigenvalue $\pm 1$):
\begin{equation}
\includegraphics[width=.9\columnwidth]{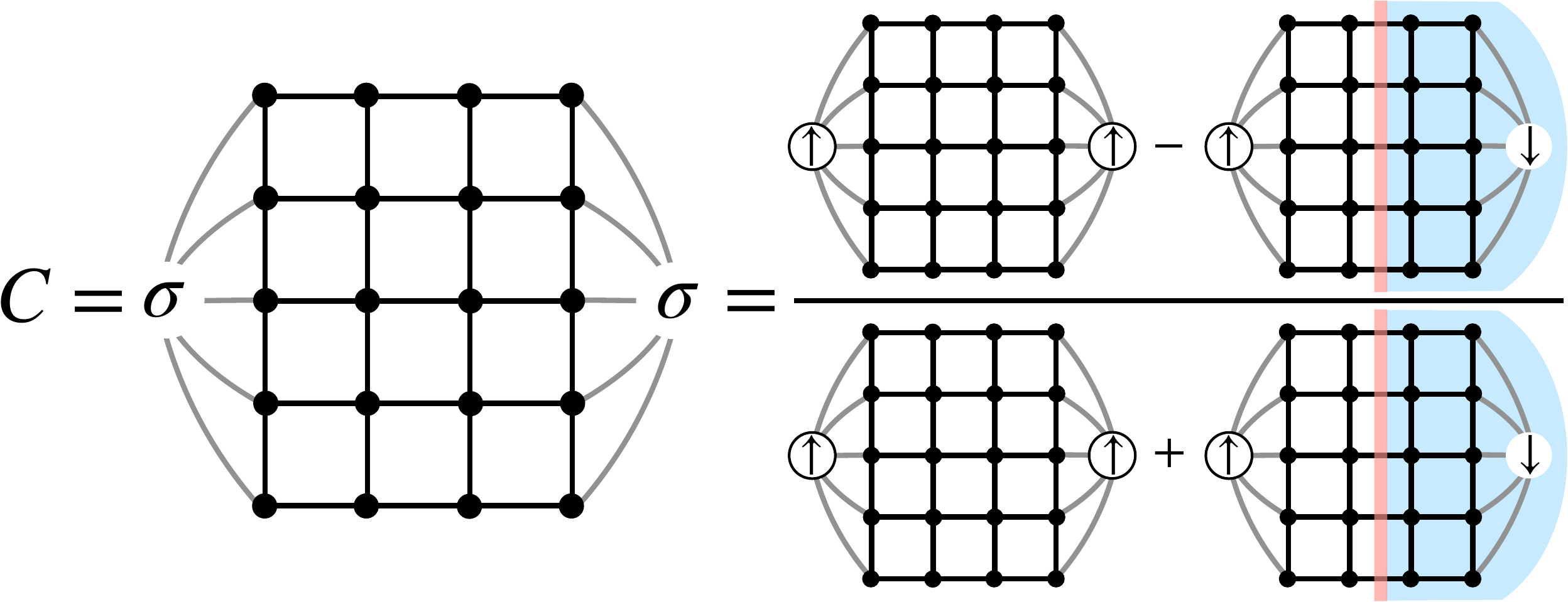} \ ,
\end{equation}
conditioned upon a given measurement record $\mathbf{s}$. Using the typical scaling dimension of the domain wall operator $\Delta_m$, we should find the typical asymptotic behavior of $C$ when $L_y\gg L_x$ for a fixed disorder realization:
\begin{equation}
   C_{\mathbf{s}} \sim  \frac{1-e^{-2\pi \Delta_m \frac{L_y}{L_x}}+\cdots}{1+e^{-2\pi \Delta_m \frac{L_y}{L_x}}+\cdots}=\tanh\left(\pi \Delta_m\frac{ L_y}{L_x}\right)+\cdots \ ,
\end{equation}
which gives a partial explanation why the domain wall entropy at the critical point does not scale with the system size, once the aspect ratio $L_y/L_x$ is fixed. Note that this is analogous to using the domain wall energy as scaling invariant probe for the Nishimori critical point at the long cylinder limit~\cite{Pujol2001,Chalker2002}.

%%%%%%%%%%%%%%%%%%%%%%%%%%%%%%%%%%%%%%%%%%%%%%%%%%%%%%%%%%%%%%%%%%%%%%%%%%%%
\subsection{Perturbative stability of the Nishimori fixed point}
%%%%%%%%%%%%%%%%%%%%%%%%%%%%%%%%%%%%%%%%%%%%%%%%%%%%%%%%%%%%%%%%%%%%%%%%%%%%
\label{NishimoriStabilityAppendix}
In this appendix, we briefly summarize the analytical argument showing the perturbative stability of the Nishimori fixed point against bond dilution, which establishes (assuming no additional intermediate fixed point) an RG flow from percolation to Nishimori. The Nishimori RG fixed point is known to be described by a CFT, with an action that we will denote $S_\star[\phi]$, with $\phi$ shorthand notation for the field degrees of freedom of the theory, with $Z= \int {\cal D} \phi \ {\rm e}^{-S_\star[\phi]}$ the corresponding partition function. Let $\epsilon(\vec{x})$ be the energy/thermal operator of the Nishimori CFT, whose scaling dimension is known numerically to be $x_1 = 1.345$. Perturbing the Nishimori CFT by this operator
\begin{equation}
S= S_\star[\phi] + \lambda \int d^2 x \epsilon(\vec{x}),
\end{equation}
drives the theory away from criticality, with an associated correlation length $\xi \sim \lambda^{-\nu}$ with $\nu = \frac{1}{2 - x_1} \simeq 1.53$. This is the exponent that we measure numerically in Fig.~\ref{fig:nu_vs_theta}. Now bond dilution amounts to making the coupling $\lambda $ space dependent and random~\cite{Cardy_1996}. To address this quenched randomness, we use a replica trick by considering $N$ replicas of the system subject to the same disorder realization
\begin{equation} 
S= \sum_{a=1}^N S_\star[\phi_a] +  \int d^2 x \lambda(\vec{x}) \epsilon_a(\vec{x}).
\end{equation}
We will assume that $\lambda$ is Gaussian distributed with zero mean and variance $\sigma^2$ (more generally, the average over disorder can be carried out using a cumulant expansion, with identical results). Averaging over disorder, we find:
\begin{equation}
S= \sum_{a} S_\star[\phi_a] - \frac{\sigma^2}{2}  \int d^2 x  \sum_{ab}\epsilon_a(\vec{x})\epsilon_b(\vec{x}).
\end{equation}
We see that bond dilution leads to an effective interaction $\epsilon_a(\vec{x})\epsilon_b(\vec{x})$ over the replicas. The terms $a=b$ simply lead to a renormalization of the critical coupling, as the (non-scaling) operator $\epsilon_a^2$ overlaps with $\epsilon_a$. Meanwhile, the terms $a \neq b$ have scaling dimension $x_{\rm dilution} = 2 x_1 \simeq 2.69$. Since $x_{\rm dilution} > 2$, we find bond dilution is {\it irrelevant} at the Nishimori fixed point. This argument is nothing but a ``Harris-like''~\cite{Harris1974,Chayes1986} inequality, $d \nu >2$ with $d=2$, which indeed holds for the Nishimori fixed point since $\nu >1$. We note that we used a replica-based field theory argument above, but this result was also established rigorously in Ref.~\onlinecite{Chayes1986} using a different approach.

%%%%%%%%%%%%%%%%%%%%%%%%%%%%%%%%%%%%%%%%%%%%%%%%%%%%%%%%%%%%%%%%%%%%%%%%%%%%
\subsection{Multifractality}
%%%%%%%%%%%%%%%%%%%%%%%%%%%%%%%%%%%%%%%%%%%%%%%%%%%%%%%%%%%%%%%%%%%%%%%%%%%%
\label{MultifractalAppendix}

In this appendix, we briefly review the multifractal nature of entanglement entropies at measurement-induced phase transitions, following Refs.~\cite{Zabalo2022,Ludwig24stat} and the earlier work in Ref.~\cite{LUDWIG1990639}.
The key observation is that the quantity
\begin{equation}
\label{LabelEqRandomVariableG}
G_n(x_1,x_2,\mathbf{s}) = \text{tr} (\rho(\mathbf{s}))^n = e^{- (n-1) S_n (\mathbf{s})},
\end{equation}
with $A=[x_1 , x_2]$ parametrizing the entanglement interval, defined for a fixed
quantum trajectory ``$s$'', behaves as a two-point correlation function in a disordered medium (due to the random nature of the measurement outcomes). More precisely, this is the two-point function of a boundary-condition changing (bcc) operator in the associated CFT~\cite{Vasseur2019,Ludwig2020}.
Upon averaging over measurement outcomes, different powers of this correlator can scale at criticality with 
exponents which are entirely independent universal numbers for different values of the moment 
order $k$~\cite{LUDWIG1990639},
\begin{equation} \label{eqMultifractalAppendix}
\overline{\left(G_n(x_1,x_2,\mathbf{s}) \right)^k} = 
\overline{e^{-k (n-1) S_n (\mathbf{s})}} 
\propto R^{-2 X_{n,k}} ,
\end{equation}
where we recall that the overbar denotes the average with respect to the Born rule probability distribution for the measurement outcomes ``$(\bf s)$''.
If this happens the scaling is referred to as ``multifractal'', and in the present context such scaling occurs for the quantity in Eq.~(\ref{eqMultifractalAppendix}).
Importantly, because the origin of this infinite hierarchy of independent critical exponents is a universal scaling form~\cite{LUDWIG1990639} of the entire probability distribution of the
random variable $G_n(x_1, x_2, {\bf s})$,
the scaling dimensions $X_{n,k}$ can be defined for 
general real indices $n$ and $k$, and thus
define a continuum of scaling dimensions. 

In practice, it is more convenient numerically to work with self-averaging quantities such as the R\'enyi entropies $S_n (\mathbf{s})$ 
 as opposed to  the quantity
 in Eq.~(\ref{LabelEqRandomVariableG}). The consequences of the multifractal scaling~\eqref{eqMultifractalAppendix} for Renyi entropies can be derived by performing a {\it cumulant expansion} of $\overline{e^{-k (n-1) S_n (\mathbf{s})}}$ and Taylor expanding $R^{-2 X_{n,k}}$
 in the moment order `$k$'. Focusing on the 2nd R\'enyi entropy ($n=2$) for simplicity, we have
\begin{equation} 
\overline{e^{-k S_2 (\mathbf{s})}}  = {\rm exp} \left(-k \kappa_1 + \frac{k^2}{2!} \kappa_2 -\frac{k^3}{3!}\kappa_3 + \dots \right), 
\end{equation}
with the cumulants defined in Eq.~\eqref{LabelEqCumulants}. Now if we Taylor expand the scaling dimension $X_{2,k} = \sum_m \frac{x_2^{(m)}}{m!} k^m $, we have
\begin{equation} 
R^{-2 X_{n,k}} =  {\rm exp} \left( -2 \left[ k x_n^{(1)} + \frac{k^2}{2!} 
x_n^{(2)} + \dots \right] \ln R \right)
\end{equation}
with $n=2$.
%%%
Comparing the latter two equations, we find in view of Eq.~(\ref{eqMultifractalAppendix})
the following scaling of all
cumulants of the 2nd R\'enyi entropy
\begin{equation} 
\kappa_m \sim 2 (-1)^{m-1} x^{(m)}_2 \ln R \,,
\end{equation}
which is the relation used in the main text. 
An analogous result is arrived at for general $n>1$. Because the density matrix
$\rho(\mathbf{s})$ appearing in Eq.~(\ref{LabelEqRandomVariableG}) is {\it normalized},
a trivial result is obtained  for the direct $n\to 1$ limit  of the moment exponents
$X_{n,k}$.

%%%%%%%%%%%%%%%%%%%%%%%%%%%%%%%%%%%%%%%%%%%%%%%%%%%%%%%%%%%%%%%%%%%%%%%%%%%%
\section{Supplementary numerical data}
%%%%%%%%%%%%%%%%%%%%%%%%%%%%%%%%%%%%%%%%%%%%%%%%%%%%%%%%%%%%%%%%%%%%%%%%%%%%

%%%%%%%%%%%%%%%%%%%%%%%%%%%%%%%%%%%%%%%%%%%%%%%%%%%%%%%%%%%%%%%%%%%%%%%%%%%%
\subsection{Computational costs}
%%%%%%%%%%%%%%%%%%%%%%%%%%%%%%%%%%%%%%%%%%%%%%%%%%%%%%%%%%%%%%%%%%%%%%%%%%%%

Our numerical codes have been run on national high-performance computing resources,
specifically the AMD EPYC (v3 Milan)-based Noctua2 cluster at the Paderborn Center for Parallel Computing (PC2),
the AMD EPYC (v4 Genoa)-based RAMSES cluster at RRZK/University of Cologne, 
and the Intel XEON Platinum 8168-based JUWELS cluster at FZ Julich.
An overview of key characteristics for production runs on these machines is provided in Table \ref{tab:hpc} above.

%%%%%%%%%%%%%%%%%%%%%%%%%%%%%%%%%%%%%%%%
\begin{figure}[b]
   \centering
   \input{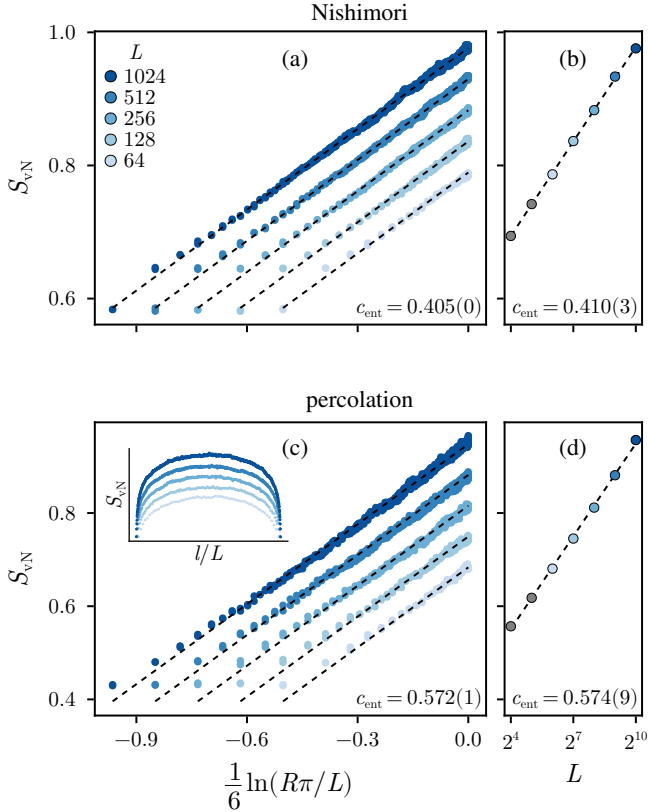}
   \caption{\textbf{Entanglement scaling: Nishimori point (top panel) versus percolation point (bottom panel).} 
      Left panel: effective central charge fit by the variation of entropy with {\it varying entanglement cut}: $S_\text{vN}(l)-S_\text{vN}(l=L/2)$ as a function of $l/L$. We fit one slope for all system sizes. 
  Right panel: the effective central charge is fit to the scaling of the {\it half-cut} entanglement entropy with the system size: $S_\text{vN}(l=L/2)$ as a function of $L$.
  For the computation a long stripe $L_y=2L_x=2L$ is chosen to guarantee the boundary 1D quantum state reaches steady state of the transfer matrix. The boundary condition is   open boundary.  
  The MPS cutoff is set to $10^{-20}$. 
    }
   \label{fig:SvN_nishi}
\end{figure}
%%%%%%%%%%%%%%%%%%%%%%%%%%%%%%%%%%%%%%%%

%%%%%%%%%%%%%%%%%%%%%%%%%%%%%%%%%%%%%%%%%%%%%%%%%%%%%%%%%%%%%%%%%%%%%%%%%%%%
\subsection{Entanglement peak location}
%%%%%%%%%%%%%%%%%%%%%%%%%%%%%%%%%%%%%%%%%%%%%%%%%%%%%%%%%%%%%%%%%%%%%%%%%%%%

One particular striking numerical result presented in the main text is the non-monotonous behavior of the effective central charge estimate $c_{\rm ent}$
along the RG flow from percolation to Nishimori shown in Fig.~\ref{fig:eff-central-charge} of the main text. Since we show data for different system sizes
$L=64,128,256,512,1024$ one might ask for a finite-size extrapolation of this data, which we present here. 

In Fig.~\ref{fig:eff-central-charge-scaling} below, we show how the peak position for increasing system sizes moves towards 
the percolation fixed point ($\theta = \pi/2$), with panel (a) showing $\theta_{\rm peak}$ as a function of inverse system size,
while panel (b) shows the distance of $\theta_{\rm peak}$ from the percolation fixed point as a function of inverse system size
in a doubly logarithmic plot. The latter data falls on top of a line, indicating a power-law fit (dashed line), indicating that peak 
position indeed moves to the percolation limit for infinite system size $L\to\infty$.
Physically, this implies a step-function behavior in the thermodynamic limit where upon moving away from the Clifford-limit of 
percolation the system immediately picks up a seizable amount of entanglement (available from spreading the Clifford-constrained
wavefunction to the fully available Hilbert space).

\begin{figure}[h!]
    \centering
    \input{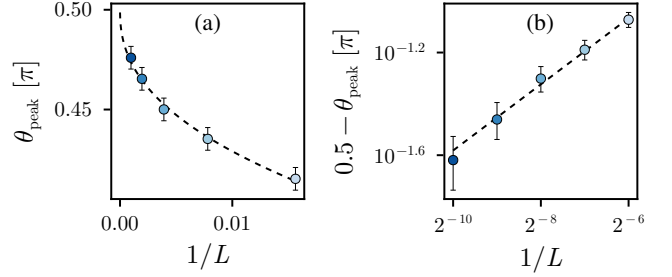}
    \caption{\textbf{Entanglement peak location}. 
    	To extrapolate the peak position of $c_{\rm ent}$ to the thermodynamic limit $L \to \infty$, 
	we show  (a) $\theta_\text{peak}$ (determined from the data in Fig.~\ref{fig:eff-central-charge}) against the inverse system size,
	and (b) its distance from $\pi/2$ versus inverse system size on a doubly-logarithmic scale. 
	We fit the power law function $f(x) = p_1 x^{p_2}$ to the data. The optimal coefficients are $p_1 = 0.509(53) \pi$ and $p_2 = 0.428(21)$.
    }
    \label{fig:eff-central-charge-scaling}
\end{figure}

%%%%%%%%%%%%%%%%%%%%%%%%%%%%%%%%%%%%%%%%%%%%%%%%%%%%%%%%%%%%%%%%%%%%%%%%%%%%
\subsection{Numerical estimation of cumulants}
\label{app:cumulants}
%%%%%%%%%%%%%%%%%%%%%%%%%%%%%%%%%%%%%%%%%%%%%%%%%%%%%%%%%%%%%%%%%%%%%%%%%%%%

\begin{figure*}[htbp]
   \centering

   \begin{minipage}[t]{\columnwidth}
      \vspace{0pt} % Align top
      \includegraphics[width=\columnwidth]{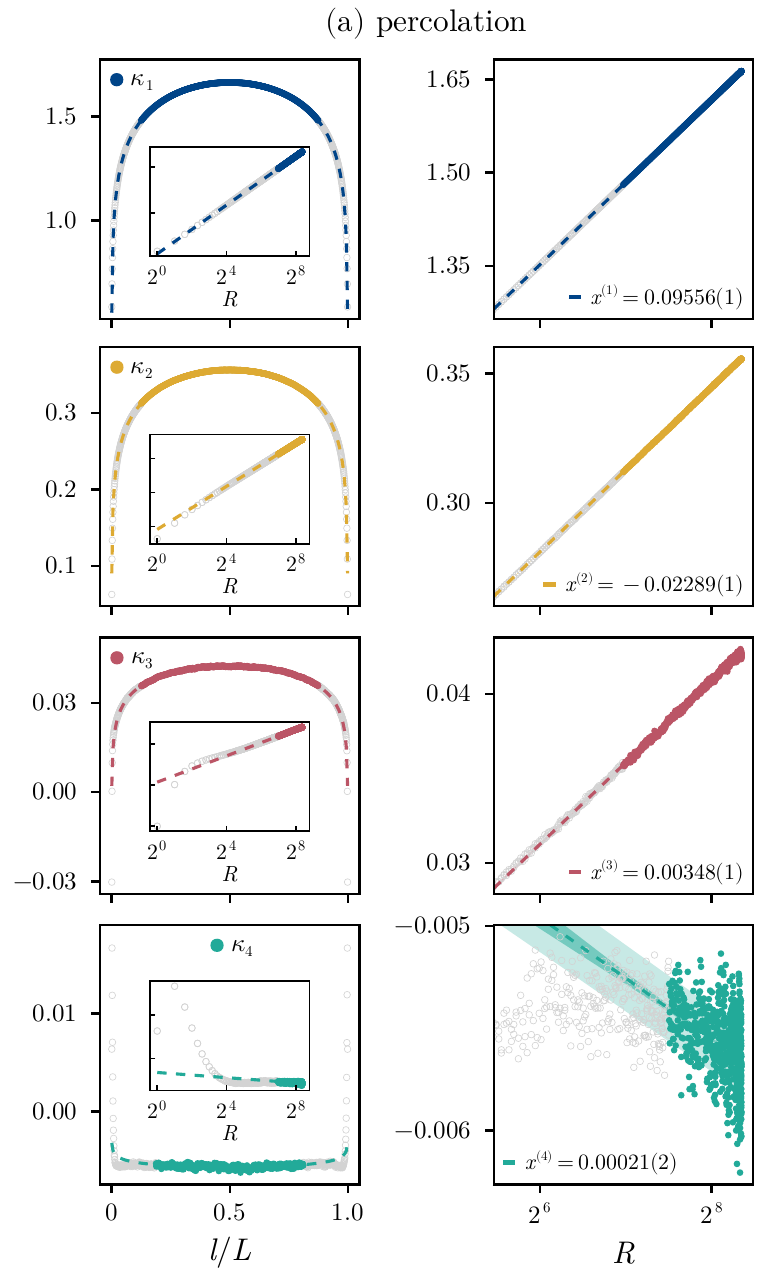}
   \end{minipage}
   \hspace{4mm}
   \begin{minipage}[t]{\columnwidth}
      \vspace{0pt} % Align top
      \includegraphics[width=\columnwidth]{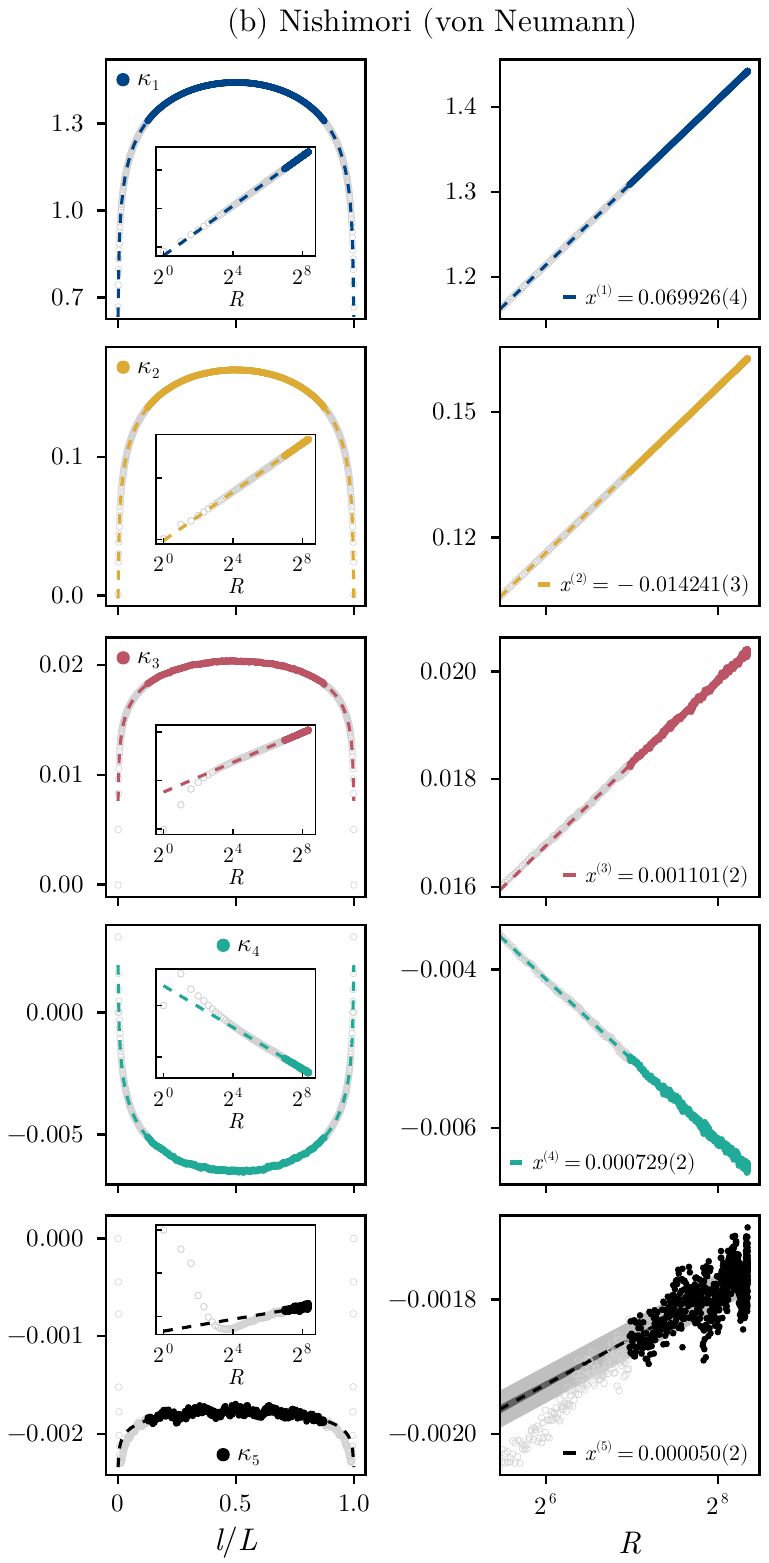}
   \end{minipage}
   \caption{\textbf{Multifractality and cumulants (periodic boundary conditions).}
      Left panels: percolation criticality.
      Right panels: Nishimori criticality.
      The underlying data is for the von Neumann entropy of system sizes $L=1024$ using {\it periodic} boundary conditions.
      Data is averaged over $\sim 14,000,000$ (percolation) and $\sim 18,000,000$ (Nishimori) samples.
      For percolation we employed Clifford simulations, while for Nishimori universality we used free fermion simulations (see main text). 
   }
   \label{fig:multifractality_percolation_nishimori_periodic_boundary}
\end{figure*}

\begin{figure*}[htbp]
   \centering
   \includegraphics[width=\columnwidth]{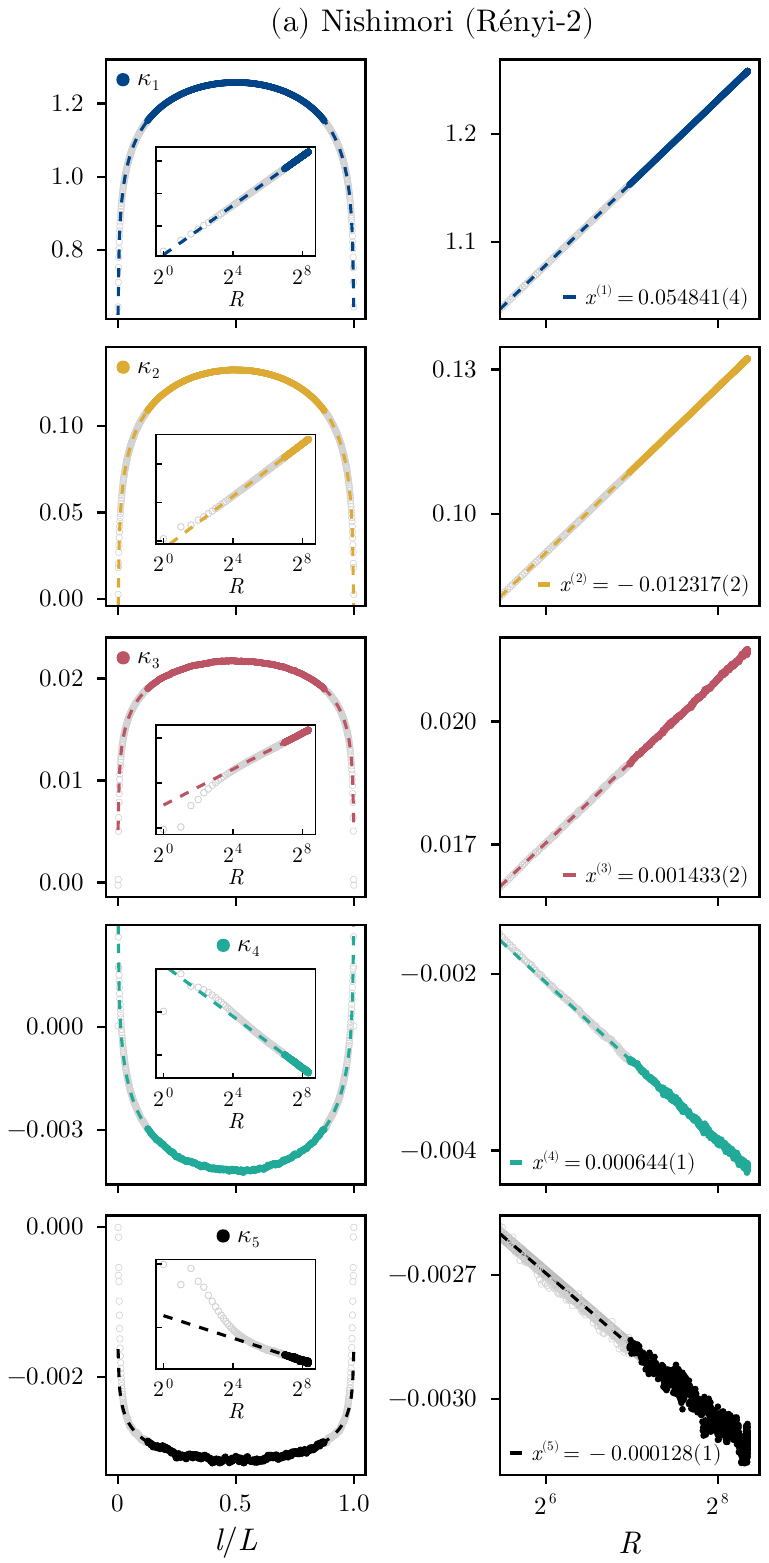}
   \hspace{4mm}
   \includegraphics[width=\columnwidth]{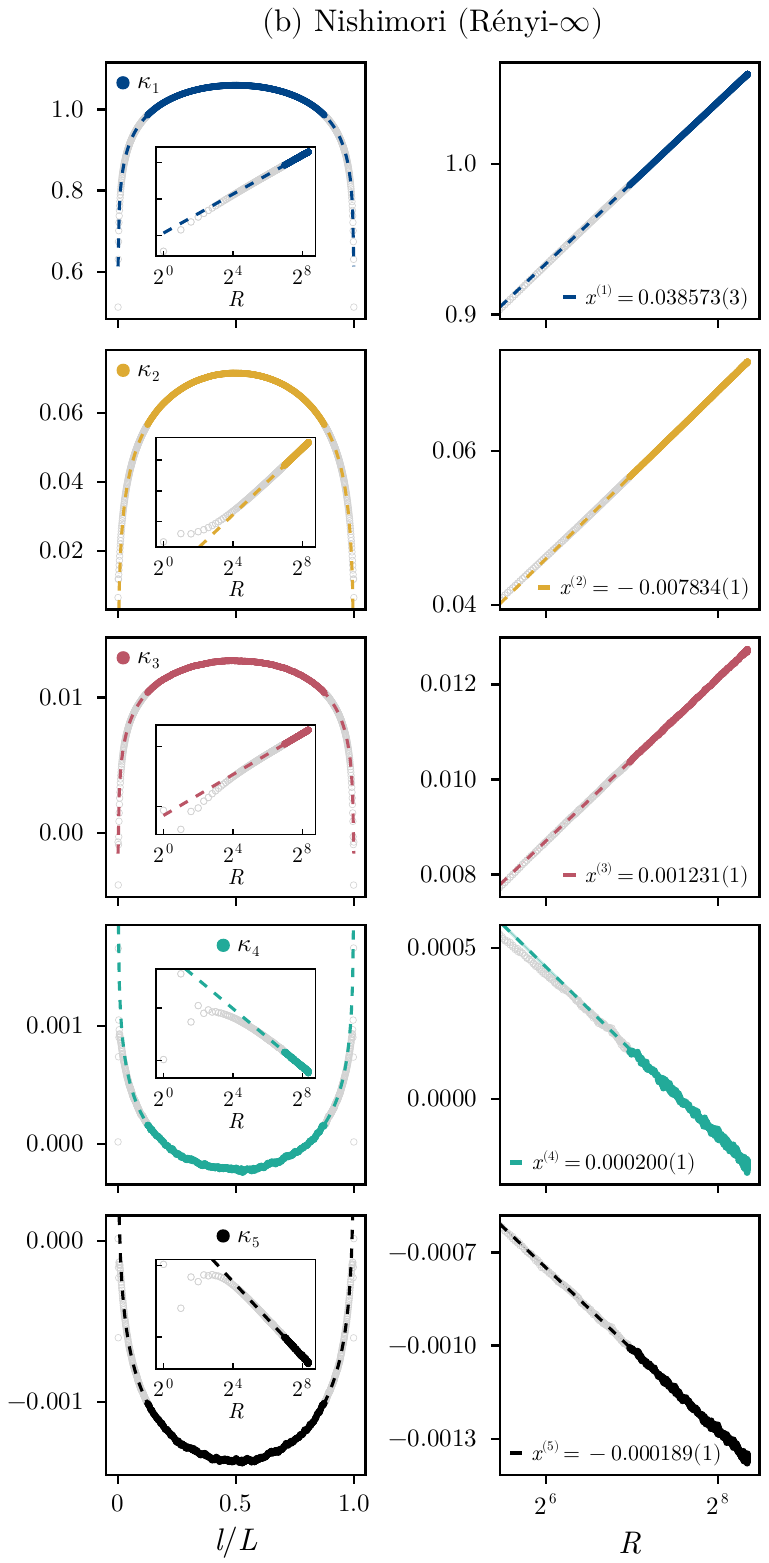}
   \caption{\textbf{Multifractality and cumulants for higher-order R\'enyi entropies (periodic boundary conditions).}
   	Same as Fig.~\ref{fig:multifractality_percolation_nishimori_periodic_boundary} 
	but for the second R\'enyi entropy (left) and $n=\infty$ R\'enyi entropy (right).
   	Data is again for system sizes $L=1024$ averaged over $\sim 18,000,000$ samples.
        Note that the fifth cumulant switches sign when going from SvN [Fig.~\ref{fig:multifractality_percolation_nishimori_periodic_boundary}(b)] 
        to the higher-order R\'enyi entropies.
   }
   \label{fig:multifractality_nishimori_Renyi}
\end{figure*}

Figs.~\ref{fig:multifractality_percolation_nishimori_periodic_boundary} and \ref{fig:multifractality_nishimori_Renyi}
provide a detailed account of our data fitting procedure to extract the leading five cumulants from fits to the entanglement entropy 
for the system sizes up to $N=1024 \times 4096$ using periodic boundary conditions.
%

%%%%%%%%%%%%%%%%%%%%%%%%%%%%%%%%%%%%%%%%%%%%%%%%%%%%%%%%%%%%%%%%%%%%%%%%%%%%
%%%%%%%%%%%%%%%%%%%%%%%%%%%%%%%%%%%%%%%%%%%%%%%%%%%%%%%%%%%%%%%%%%%%%%%%%%%%

%%%%%%%%%%%%%%%%%%%%%%%%%%%%%%%%%%%%%%%%%%%%%%%%%%%%%%%%%%%%%%%%%%%%%%%%%%%%
\subsubsection*{Nishimori versus percolation criticality}
%%%%%%%%%%%%%%%%%%%%%%%%%%%%%%%%%%%%%%%%%%%%%%%%%%%%%%%%%%%%%%%%%%%%%%%%%%%%

Concentrating first on the data in Fig.~\ref{fig:multifractality_percolation_nishimori_periodic_boundary} one can see that the Nishimori
universality class leads to cleaner data than percolation criticality, which one might rationalize by the fact that the percolation/Clifford problem 
is obtained from averaging {\it discrete} data, which requires more samples to converge.

For both universality classes, the ``arc structure" of the entanglement entropy, when plotted as a function of subsystem size $l/L$, is clearly visible 
(as demanded by the analytical form of Eq.~\ref{UnitaryCFTEntropy}) and switches its overall sign for the fourth-order cumulant (as expected analytically
for percolation, but unknown in the case of Nishimori universality).

When trying to fit the fourth order cumulant, we note that this is a challenging task for the percolation data. In fact, there is a strong dependence of the
numerical estimate of the cumulant when restricting the data set to only fit the bulk data in the middle of the arc. What an ideal choice might look like is
not at all obvious and we have ``cherry-picked" the interval to obtain an estimate close to the analytical prediction (which, of course, does not serve as
an independent confirmation anymore). 
In contrast, there is no such need for outside guidance for the case of Nishimori criticality where we can provide a high-precision estimate for the 
fourth-order cumulant from following the same fitting protocol as for all other cumulants. We can even expand this approach to obtain an estimate
for the fifth-order cumulant, as shown in the lower right panel of Fig.~\ref{fig:multifractality_percolation_nishimori_periodic_boundary}, though the fit is clearly of much lesser quality than the other cumulants.

%%%%%%%%%%%%%%%%%%%%%%%%%%%%%%%%%%%%%%%%%%%%%%%%%%%%%%%%%%%%%%%%%%%%%%%%%%%%
%%%%%%%%%%%%%%%%%%%%%%%%%%%%%%%%%%%%%%%%%%%%%%%%%%%%%%%%%%%%%%%%%%%%%%%%%%%%

%%%%%%%%%%%%%%%%%%%%%%%%%%%%%%%%%%%%%%%%%%%%%%%%%%%%%%%%%%%%%%%%%%%%%%%%%%%%
\subsubsection*{Higher-order R\'enyi entropies}
%%%%%%%%%%%%%%%%%%%%%%%%%%%%%%%%%%%%%%%%%%%%%%%%%%%%%%%%%%%%%%%%%%%%%%%%%%%%

Notably, the quality of the cumulant fits further improves as one goes from the $n=1$ von Neumann entropy to higher-order R\'enyi entropies,
with a direct comparison shown in Fig.~\ref{fig:multifractality_nishimori_Renyi} for the second and $n=\infty$ R\'enyi entropies. 
Of particular interest might be the behavior of the fifth-order cumulant, which exhibits a sign change when going from the first to second 
R\'enyi entropy (and then remains negative for all higher-order R\'enyi entropies).

%%%%%%%%%%%%%%%%%%%%%%%%%%%%%%%%%%%%%%%%%%%%%%%%%%%%%%%%%%%%%%%%%%%%%%%%%%%%
\subsubsection*{Boundary conditions}
%%%%%%%%%%%%%%%%%%%%%%%%%%%%%%%%%%%%%%%%%%%%%%%%%%%%%%%%%%%%%%%%%%%%%%%%%%%%

\begin{figure*}[th!]
   \centering
   \includegraphics[width=\textwidth]{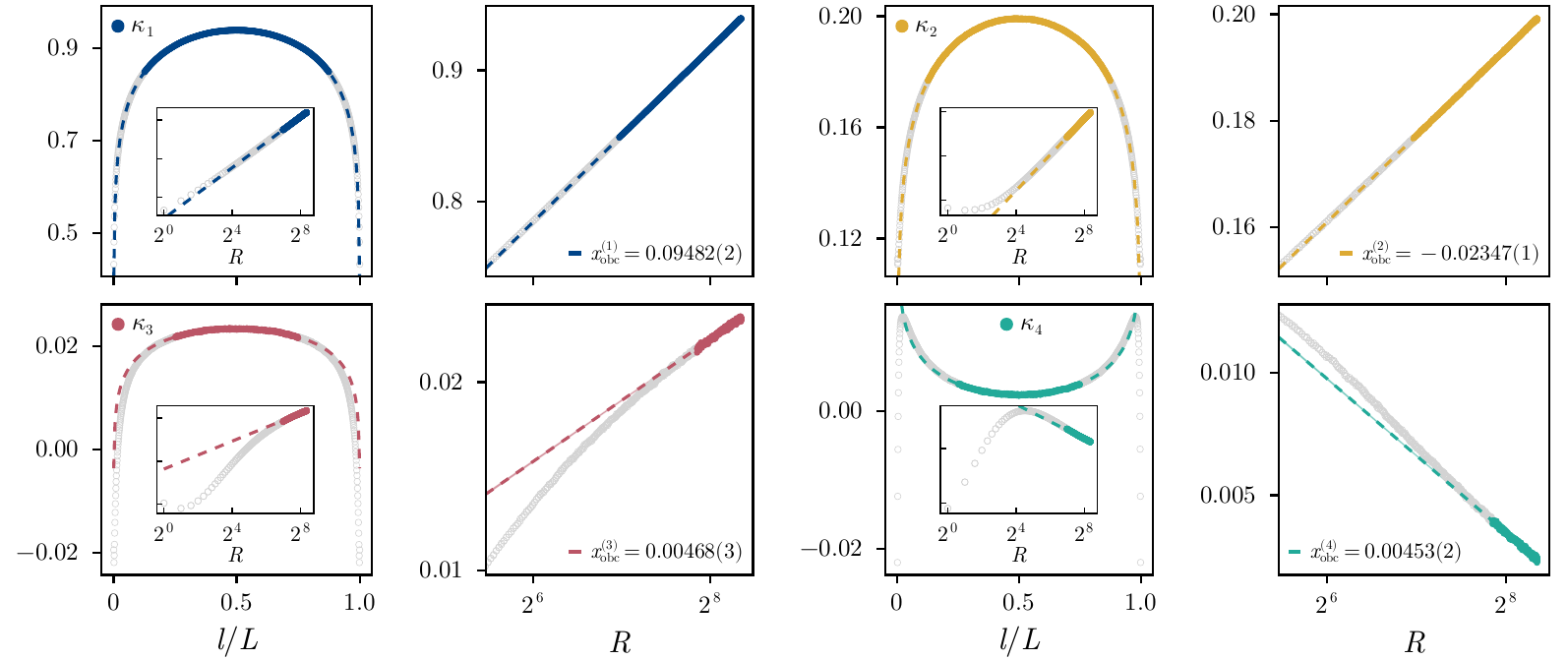} 
   \caption{ {\bf Multifractality and cumulants for percolation (open boundary conditions).} 
      The underlying data is for the von Neumann entropy of system sizes $L=1024$ using {\it open} boundary conditions. 
      Data is calculated using Clifford simulations and averaged over $\sim 14,000,000$ samples. 
      Note that, starting from the third cumulant, we observe boundary effects that further increase for the fourth cumulant,
      cf.\ Fig.~\ref{fig:multifractality_percolation_nishimori_periodic_boundary}(a) where such effects are absent for {\it periodic} boundary conditions.
      To minimize such boundary effects on third and fourth cumulant estimates we only fit half of the data points in the bulk. 
   }
    \label{fig:multifractality_percolation_open_boundary}
\end{figure*}

Let us close with the observation that, in our simulations, {\it periodic} boundary conditions 
(see, e.g., Fig.~\ref{fig:multifractality_percolation_nishimori_periodic_boundary}) lead to much cleaner
data than {\it open} boundary conditions (Fig.~\ref{fig:multifractality_percolation_open_boundary}). 
This might be a useful pointer for any future calculations of cumulant estimators.

%%%%%%%%%%%%%%%%%%%%%%%%%%%%%%%%%%%%%%%%%%%%%%%%%%%%%%%%%%%%%%%%%%%%%%%%%%%%
%%%%%%%%%%%%%%%%%%%%%%%%%%%%%%%%%%%%%%%%%%%%%%%%%%%%%%%%%%%%%%%%%%%%%%%%%%%%

\end{document}